%% file: paper.tex
\documentclass{template}

\usepackage[dcucite]{harvard}
\usepackage{amssymb,amsfonts,latexsym,amsmath}
\usepackage{comment}
\usepackage{multirow}
\usepackage{enumerate}
\usepackage{wrapfig}
\usepackage[dvips]{epsfig}
\usepackage[english]{babel}

\newenvironment{example}{\paragraph{Example}}{\hfill $\Box$ \bigskip}

\begin{document}
\label{firstpage}

\title{Compressed Vertical Partitioning for Full-In-Memory 
	RDF Management}
\author[S. \'Alvarez-Garc\'ia et al.]
{	Sandra \'Alvarez-Garc\'ia$^1$, Nieves Brisaboa$^1$, 
	Javier D. Fern\'andez$^{2,3}$, \\
	{\large Miguel A. Mart\'inez-Prieto$^{2,3}$, and Gonzalo Navarro$^3$}
	\vspace{0.2cm} \\ 
$^1$ Database Lab, Facultade de Inform\'atica, University of A Coru\~na, Spain \\ 
$^2$ DataWeb Research, Department of Computer Science, University of Valladolid, Spain \\
$^3$ Department of Computer Science, University of Chile, Chile}

\maketitle

\begin{abstract}
The Web of Data has been gaining momentum in recent years. This leads to 
increasingly publish more and more semi-structured datasets following, in many
cases, the RDF data model based on atomic {\em triple} units of {\em subject},
{\em predicate}, and {\em object}. Although it is a very simple model,
specific compression methods become necessary because datasets are 
increasingly larger and various scalability issues arise around their
organization 
and storage. This requirement is even more restrictive in RDF stores because 
efficient SPARQL resolution on the compressed RDF datasets is also 
required.

This article introduces a novel RDF indexing
technique that supports efficient SPARQL resolution in compressed
space. Our technique, called k$^2$-{\sc triples}, uses the predicate to 
vertically partition the dataset into disjoint subsets of pairs (subject, 
object), one per predicate. These subsets are represented as binary matrices of subjects 
$\times$ objects in which 1-bits mean that the corresponding triple exists in
the dataset. This model results in very sparse matrices, which
are efficiently compressed using k$^2$-trees. We enhance this model with
two compact indexes listing the predicates related to each different subject
and object in the dataset, in order to address the specific weaknesses of
vertically partitioned representations. The resulting technique not only 
achieves by far the most compressed representations, but also achieves the best 
overall performance for RDF retrieval in our experimental setup. 
Our approach uses up
to $10$ times less space than a state of the art baseline, and outperforms
its time performance by several order of magnitude on the most basic query 
patterns. 
In addition, we optimize traditional join algorithms on 
k$^2$-{\sc triples} and define a novel one leveraging its specific features. 
Our experimental results show that our technique also overcomes
traditional vertical partitioning for join resolution, reporting the best numbers
for joins in which the non-joined nodes are provided, and being competitive in
the great majority of the cases.
\end{abstract}


\input{intro}

\input{state}
\input{cds}
\input{k2triples}

\input{experimentation}
\input{conclusions}

\begin{acknowledgements}
This work was partially funded by the Spanish Ministry of Economy and Competitiveness, projects TIN2009-14560-C03-02 (first and second authors) and TIN2009-14009-C02-0 (third and fourth authors), the Xunta de Galicia with FEDER refs. 2010/17, CN2012/211 (first and second authors), and Chilean Fondecyt 1-110066 (fourth and fifth authors). The first author is granted by the Spanish Ministry of Economy and Competitiveness ref. BES-2010-039022. The third author is granted by the Regional Government of Castilla y Leon (Spain) and the European Social Fund. The fourth author has a Ibero-American Young Teachers and Researchers Grant funded by Santander Universidades. \\
\end{acknowledgements}	
\vspace{-0.5cm}
	
	\bibliographystyle{agsm}
	\bibliography{paper}

\label{lastpage}

\end{document}

%% file: intro.tex
\section{Introduction}
\label{s:intro}

The {\em Resource Description Framework} (RDF) \cite{RDF:04} provides a simple
scheme for structuring and linking data that describe facts of the world 
\cite{BHBL:09}. It models knowledge in the form of triples {\tt (subject, 
predicate, object)}, in which the {\em subject} represents the resource being
described, the {\em predicate} is the property, and the {\em object} contains 
the value associated to the property for the given subject. RDF was originally
conceived (under a {\em document-centric} perspective of the Web) as a 
foundation for processing metadata and describing resources. However, this 
conception does not address its actual usage. The current 
Recommendation\footnote{\tt http://www.w3.org/TR/rdf-syntax-grammar/}
considers RDF as the key to do for machine processable information (application
data) what the WWW has done for hypertext, that is, to allow data to be 
processed outside the particular environment in which it was created, in a 
fashion that can work at Internet scale. This statement describes
the RDF status in the so-called {\em Web of Data}.

The Web of Data materializes the basic principles of the {\em Semantic Web} 
\cite{BLHL:01} and interconnects datasets from diverse fields of knowledge 
within a cloud of data-to-data hyperlinks that enables a ubiquitious and 
seamless data integration to the lowest level of granularity. Thus, information
follows a {\em data-centric} organization within the Web of Data. The advancement
in extraction mechanisms \cite{GGHD:13} or the annotation of massive amounts of
resources \cite{SIM:11}, among others, have motivated the growth of the Web of Data
in which the number (and scale) of semantic 
applications in use increases, more RDF data are linked together, and increasingly 
larger datasets are obtained. This popularity is the basis for the development of RDF 
management systems (referred to as {\em RDF stores}), which play a central
role in the Web of Data. They provide support for RDF storage and also lookup 
infrastructure to access it via SPARQL \cite{SPARQL:08} query interfaces. 
Although the increased amount of RDF available is good for semantic 
initiatives, it is also causing performance bottlenecks in the RDF stores
currently used \cite{HAR:11}. Thus, scalability arises as a major issue,
restricting popular RDF applications, like inference-based ones, because
traditional solutions are not suitable for large-scale deployments \cite{JJB:09}.
The scalability management is closely related to the RDF physical organization, 
storage and the mechanisms designed for its retrieval. 

Two families of RDF stores are mainly used within the current Web of Data. On
the one hand, {\em relational} stores are built on the generality, scalability,
and maturity of relational databases. Some logical models have been proposed 
for organizing and storing RDF in relational stores \cite{SAN:10}. However, the
relational model is quite strict to fit semi-structured RDF features, and 
these systems reach only limited success. On the other hand, {\em native} 
solutions are custom designed from scratch and focus directly on specific RDF
features. Although various techniques have been proposed, multi-indexing
ones are the most used within the current state of the art \cite{WKB:08,ACZH:10,NW:10}. 
Even so, these approaches also suffer from lack of scalability because they
raise huge space requirements. %% and pay many I/O costs in query resolution. 

We address this scenario with one main guideline: {\em to reduce 
the space required for organizing and storing RDF}. Spatial savings reduce 
storage space, but also have significative impact in processing times because
they allow us to represent more data in the same space. This fact 
enables that larger datasets can be managed and queried in main memory, so that
I/O costs are reduced and querying processes are completed earlier. Our 
approach, called {k$^2$-{\sc triples}, leverages this fact to manage larger RDF
datasets in main memory.

k$^2$-{\sc triples} revisits vertical partitioning \cite{AMMH:07} by replacing
relational storage by compact k$^2$-tree stuctures \cite{BLN:09,L:11}. The 
k$^2$-tree provides indexed access to binary matrices and excels in compression
when these matrices are very sparse. This case arises when an RDF dataset is
vertically partitioned, because the number of subjects related to pairs 
object-predicate, and the number of objects related to pairs subject-predicate,
are very few for real-world datasets \cite{FMPGPA:13}. This fact not only yields
a space effective approach, outperforming the space of the best state of the art
alternatives by a factor of 1.5 to 12. The k$^2$-tree representation also enables 
efficient RDF retrieval for triple patterns, which are the basic SPARQL queries.
Our representation is up to 5--7 orders of magnitude faster than the state of 
the art to resolve most triple patterns.

Our basic k$^2$-{\sc triples} approach is further enhanced with additional
compact data structures to speed up the processing of advanced SPARQL queries,
in particular those with no fixed predicate.
This is the main weakness of
vertical partitioning and is directly related to the number of different 
predicates used for modeling a dataset. We define two compact indexes that
list the predicates related to each subject and each object
in the dataset. These structures involve an acceptable space overhead 
(less than the $30\%$ of the original space requirements), and improves 
performance by more than an order of magnitude when these classes of queries
are performed on a dataset comprising a large predicate set.

We also focus on join operations, because these are the basis for 
building the Basic Graph Patterns (BGPs) commonly used in SPARQL. We optimize 
the traditional {\em merge} and {\em index} join algorithms to take advantage of
the basic retrieval functionalitiy provided in k$^2$-{\sc triples}. Besides, 
we describe an innovative join algorithm that traverses several k$^2$-trees in 
parallel and reports excellent results in many practical scenarios. 
Our technique sharply overcomes traditional vertical partitioning
in join resolution, reporting up to $5$ orders of magnitude of improvement in
joins involving any variable predicate. The comparison with more advanced
techniques shows that k$^2$-{\sc triples} performs up to 3 orders of magnitude
faster in joins providing the non-joined nodes, while remaining competitive in the other
ones.

Our experiments compare  k$^2$-{\sc triples} with various state of the art
alternatives on various real-life RDF datasets, considering space and query 
time. In summary, k$^2$-{\sc triples} achieves the most compressed RDF 
representations to the best of our knowledge, representing $\approx 200,000$ 
triples/MB in our largest dataset ({\tt dbpedia}),
 where the next best techniques in
space usage ({\tt MonetDB} and {\tt RDF-3X}, in this order)
represent $125,000$ and $25,000$ triples per MB. When solving basic triple
patterns, our enhanced structure requires 0.01 to 1 millisecond (msec) per query
on {\tt dbpedia}, whereas the next fastest alternative ({\tt RDF-3X}) takes
1 to 10 msec and {\tt MonetDB} takes 100 msec to 1 second.
Finally, our best numbers in join resolution range from 0.01 
to 10 msec per query (also on {\tt dbpedia}), whereas RDF-3X always requires
over 10 msec and MonetDB wastes more than 1000 seconds in the same cases.
Generally our times are below 0.1 seconds per query, which is comparable
to the best peformance reported in the state of the art using RDF3x.

The paper is organized as follows. The next section gives basic notions about
RDF and SPARQL, and reviews the state of the art on RDF stores. Section 
\ref{s:cds} introduces compact data structures and details the k$^2$-tree index 
used as the basis 
for our approach. The next 
three sections give a full description of our approach: Section \ref{s:k2t1} 
explains how k$^2$-trees are used for physical RDF organization and storage, 
Section \ref{s:k2t2} describes the mechanisms designed for basic RDF retrieval
over this data partitioning, and Section \ref{s:k2t3} details the join 
algorithms designed as the basis for BGP resolution in SPARQL. 
Section \ref{s:exper} experimentally compares k$^2$-{\sc triples} with 
state of the art RDF stores on various real-world 
datasets, focusing both in space usage and query time performance. 
Finally, Section \ref{s:conc} gives our conclusions about the current scope of 
{k$^2$-{\sc triples}} and devises its future evolution within the Web of Data.

%% file: state.tex
\section{State of the Art}
\label{s:state}

The marriage of RDF and SPARQL is a cornerstone of the Web of Data because they
are the standards recommended by the W3C for describing and 
querying semantic data. Both are briefly introduced to give basic 
notions about their use and properties.

As previously described, {\bf RDF} \cite{RDF:04} provides a description
framework for structuring and linking data in the form of {\em triples}
{\tt (subject, predicate, object)}. A triple can also be seen as an edge in
labeled
graph, ${\tt S}\xrightarrow{\tt P} {\tt O}$, where the subject {\tt S} and
the object {\tt O} are represented as vertices and the predicate {\tt P} labels
the edge that joins them. The graph modeling the whole triple set is called
the RDF Graph, a term
formally introduced in the RDF Recommendation \cite{RDF:04}.

\begin{figure}
  \centering
  \includegraphics*[bb=18 372 577 493, scale=0.6]{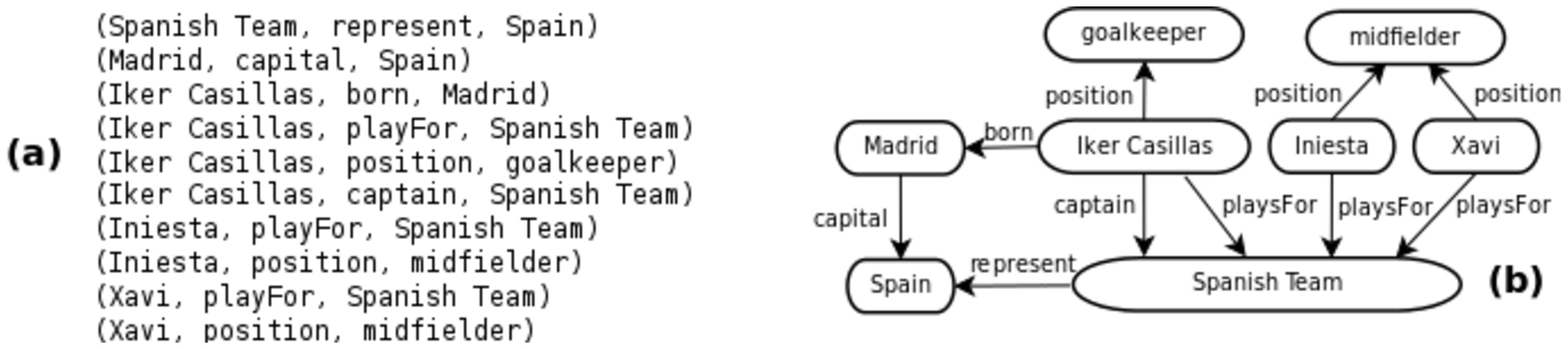}
  \caption{\label{fig:exrdf} Example of RDF-based modelling.}
  \vspace{0.5cm}
  \includegraphics*[bb=18 395 577 468, scale=0.6]{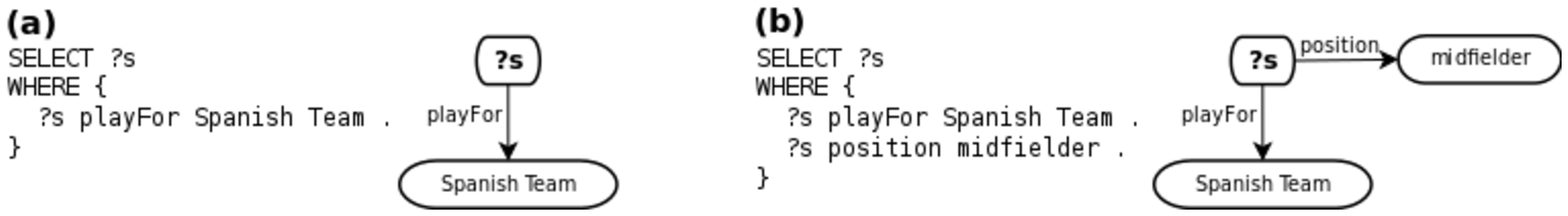}
  \caption{\label{fig:exsparql} Examples of (a) SPARQL triple pattern and 
    (b) SPARQL basic graph pattern.}
\end{figure}

Figure \ref{fig:exrdf} models in RDF some information related to the 
{\em Spanish National Football Team} (hereinafter referred to as {\tt Spanish 
Team}) and some of its players. Two equivalent notations are considered: (a) 
enumerates the set of triples representing this information, whereas (b) shows
its equivalent graph-based representation. Following the triples representation
in (a), we firstly state that the {\tt Spanish Team} {\tt represent}s 
{\tt Spain} and {\tt Madrid} is the {\tt capital} of {\tt Spain}. Then, we 
describe the player {\tt Iker Casillas}: he was {\tt born} in {\tt Madrid}, 
{\tt play}s {\tt for} the {\tt Spanish Team} in the {\tt position} of 
{\tt goalkeeper} and he is also the team {\tt captain}. Finally, both 
{\tt Iniesta} and {\tt Xavi} {\tt play for} the {\tt Spanish Team} in the 
{\tt position} of {\tt midfielder}. These same relations can be found by 
traversing the labelled edges in the graph (b).

{\bf SPARQL} \cite{SPARQL:08} is the W3C Recommendation for querying RDF. It is
a graph-matching language built on top of {\em triple patterns}, that is, RDF 
triples in which each subject, predicate or object may be a variable. This 
means, that eight different triple patterns are possible in SPARQL (variables
are preceded, in the pattern, by the symbol {\tt ?}): {\tt (S,P,O), (S,?P,O),
(S,P,?O), (S,?P,?O), (?S,P,O), (?S,?P,O), (?S,P,?O)}, and {\tt (?S,?P,?O)}.

%% \begin{itemize}
%%   \item{\tt (S,P,O)}: checks if the triple {\tt (S,P,O)} is or not in the 
%%     RDF dataset.
%%   \item{\tt (S,?P,O)}: returns all {\em predicates} relating the given pair of 
%%     subject and object {\tt (S,O)}.
%%   \item{\tt (S,P,?O)}: returns all {\em objects} related to the subject {\tt S}
%%     through the predicate {\tt P}.
%%   \item{\tt (S,?P,?O)}: returns all pairs {\em (predicate, object)} associated 
%%     to the given subject {\tt S}.
%%   \item{\tt (?S,P,O)}: returns all subjects related to the object {\tt O} 
%%     through the predicate {\tt P}.
%%   \item{\tt (?S,?P,O)}: returns all pairs {\em (subject, predicate)} associated
%%     to the given object {\tt O}.
%%   \item{\tt (?S,P,?O)}: returns all pairs {\em (subject, object)} associated 
%%     through the predicate {\tt P}.
%%   \item{\tt (?S,?P,?O)}: returns all triples in the RDF dataset.
%% \end{itemize}

SPARQL builds more complex queries (generically referred to as {\em Basic Graph
Patterns}: BGPs) by joining sets of triple patterns. Thus, competitive SPARQL 
engines require, at least, fast triple pattern resolution and efficient join
methods. Additionally, query optimizers are required to build efficient 
execution plans that minimize the amount of intermediate results to be joined
in the BGP. Query optimization is orthogonal to the current work, and any 
existing technique can be implemented on top of {k$^2$-{\sc triples}}.

Figure \ref{fig:exsparql} shows two simple SPARQL queries over the RDF excerpt
described in the example above:

\begin{itemize}
  \item The first query (left), expressed in normative SPARQL syntax on the 
    left of the figure, represents the triple pattern {\tt (?S,P,O)}. It asks 
    for all subjects related to the {\tt Spanish Team} through the predicate
    {\tt playFor}. From a structural perspective (bottom), this query is a 
    subgraph comprising two nodes connected through the edge labeled 
    {\tt playFor}: the destination node represents the element {\tt Spanish
    Team}, whereas the source node is a variable. This way, the query resolution
    involves graph pattern matching for locating all nodes that can play the 
    source role in this query subgraph. In this case, the valid nodes represent
    the players {\tt Iker Casillas}, {\tt Iniesta}, and {\tt Xavi}. 
  \item The second query restricts the previous one for only retrieving the
    {\tt midfielder} players of the {\tt Spanish Team}. This refinement is 
    implemented through a second triple pattern {\tt (?S,P,O)} setting the 
    predicate {\tt position} and the object {\tt midfielder}. As can be seen on
    the right figure, the two triple patterns of the query are joined by their 
    subject. Its resolution matches the query subgraph to the RDF graph, and 
    retrieves the elements conforming to the variable node; in this case, the 
    result contains the players {\tt Iniesta} and {\tt Xavi}.
\end{itemize}

RDF is defined as a logical data model, so no restrictions are posed on its 
physical representation and/or storage. However, its implementation has a clear
effect on the retrieval efficiency, and therefore on the success of a SPARQL
engine within an RDF store. We review below the existing techniques for 
modeling, partitioning, and indexing RDF, and discuss their use in some real
RDF stores. Our goal is to show the achievements and shortcomings in 
the state of the art to highlight the potential for improvement on which our 
work focuses. We firslty show the approaches based on a relational 
infrastructure, and then the solutions natively designed for handling RDF.

\subsection{Relational Solutions}

Some logical schemes have been proposed for representing RDF over the 
infrastructure provided by relational databases, but their success has been
limited due to the strictness of the relational model for handling the 
semi-structured RDF features. Nevertheless, there is still room for 
optimization in the field of relational solutions for RDF management 
\cite{SAN:10}, and we describe below the most used schemes.

\paragraph{Single three-column table} This is the most straightforward 
scheme for modelling RDF over relational infrastructure. It represents RDF
triples as generic tuples in a huge three-column table, so generic BGP
resolution involves many expensive self-joins on this table. Systems such as
3store \cite{HG:03} or the popular
Virtuoso\footnote{\scriptsize\tt http://www.openlinksw.com/} implement this
scheme.

\paragraph{Property tables} This model arises as a more practical scheme 
for RDF organization in relational databases because it proposes to create
relational-like property tables out of RDF data. These tables gather together
information about multiple predicates (properties) over a list of subjects.
Thus, a given property table has many columns as different predicates (one
per column) are used
for describing the subjects that it stores (in rows). Although this model reduces 
significantly the number of self-joins, the cost of the query resolution 
remains high. Besides, the use of property tables induces two additional 
problems. On the one hand, storage requirements increase because NULL values
must be explicitly stored, in each tuple, if the represented subject is not 
described for a given property in the table. On the other hand, multi-valued 
attributes are abundant in semantic datasets and they are somewhat awkward to
express in property tables \cite{AMMH:07}. Thus, property tables are a 
competitive choice for representing well-structured datasets, but they lose 
potential in a general case. Systems like Jena \cite{W:06} or Sesame 
\cite{BKH:03} use property tables for modeling RDF.

\paragraph{Vertical partitioning} The vertical partitioning (VP) scheme 
\cite{AMMH:07} can be seen as a specialized case of property tables in which
each table gathers information about a single predicate. This way, VP uses
many tables as different predicates are used in the dataset, each one 
storing tuples {\tt (S,O)} that represent all (subject,object) pairs related 
through a given predicate. Each table is sorted by the subject column, in 
general, so particular subjects can be located quickly, and fast merge joins 
can be used to reconstruct information about multiple properties for subsets
of subjects \cite{AMMH:07}. However, this decision penalizes queries by object
that require additional object indexing for achieving competitive 
resolution.

VP-based solutions avoid the weaknesses previously reported for property tables
because only non-NULL values are stored, and multi-valued attributes are listed
as successive tuples in the corresponding table. However, VP-based solutions 
suffer from an important lack of efficiency for resolving queries with unbounded 
predicate; in this case, all tables must firstly be queried and their results 
must then be merged to obtain the final result. This cost increases linearly 
with the number of different predicates used in the dataset, so VP is not the
best choice for representing datasets with many predicates.

Abadi, {\em et al.} \cite{AMMH:07,AMMH:09} report that querying performance in
column-oriented databases is up to one order of magnitude better than that 
obtained in row-oriented ones. This fact motivates the implementation of their
system SW-Store as an extension of the column-oriented database C-Store 
\cite{SABal:05}. SW-Store leverages all the advantages reported above, but also
suffers from a lack of scalability for queries with unbounded predicate. SW-Store, 
like some other approaches (such as the reviewed below: Hexastore, RDF3X, and 
BitMat), firstly perform a dictionary encoding that maps long URIs and literal
values to integer IDs. This decision allows triples to be rewritten as three-ID
groups, and this is the representation finally stored in the database. 
%% In addition to the VP scheme, SW-Store also indexes some 
%% materialized path expressions. This speeds up path expressions resolution at 
%% the price of increasing storage requirements. 
Sidirourgos, {\em et al.} \cite{SGKNM:08} show additional experiments on VP. 
They replace C-Store by MonetDB\footnote{\scriptsize\tt http://www.monetdb.org/}
in the database layer; these systems show a couple of differences 
\cite{SHKLP:08}: i) data processing in C-Store is disk-based while it is 
memory-based in MonetDB; and ii) C-Store implements carefully optimized merge 
joins and makes heavy use of them, whereas MonetDB uses merge joins less 
frequently. Even so, MonetDB arises as a competitive choice in this scenario 
\cite{SGKNM:08}. 
%% The findings reported in these works differ from each. Whereas Abadi, {\em et 
%% al.} \cite{AMMH:07,AMMH:09} conclude that VP overcomes property tables, 
%% Sidirourgos, {\em et al.} \cite{SGKNM:08} refutes this conclusion and show that
%% the comparison depends on the dataset features. 

\subsection{Native Solutions}

Native solutions are custom designed from scratch to better address RDF 
peculiarities. Although some works \cite{BHS:03,HGu:03,AG:05} propose
different graph-based models, the main line of research focuses on 
{\bf multi-indexing} solutions. Harth and Decker \cite{HD:05} propose a 
six-index structure for managing {\em quads}\footnote{A {\em quad} can be
regarded as a triple enhanced with a fourth component of {\em provenance}:
{\tt (s,p,o,c)}, where {\tt c} is the context of the triple {\tt (s,p,o)}.}.
This scheme allows all quads conforming to a given query pattern (in which 
the context can also be a variable) to be quickly retrieved. This experience
has been integrated in some systems within the current state of the art for 
RDF management. 

\paragraph{Hexastore}\cite{WKB:08} It adopts the rationale of
VP and multi-indexing, but takes it further, to its logical conclusion. In
contrast to VP, Hexastore treats subjects, predicates, and objects equally.
That is, whereas VP prioritizes predicates and indexes pairs (subject,object)
around them, Hexastore builds specific indexes around each dimension and defines
a prioritization between the other two:

\begin{itemize}
  \item For each subject {\tt S}, two representations {\tt (P,O)} and 
    {\tt (O,P)} are built.
  \item For each predicate {\tt P}, two representations {\tt (S,O)} and
    {\tt (O,S)} are built.
  \item For each object {\tt O}, two representations {\tt (S,P)} and 
    {\tt (P,S)} are built.
\end{itemize}

This way, Hexastore manages six indexes: {\tt (S,P,O)}, {\tt (S,O,P)}, 
{\tt (P,S,O)}, {\tt (P,O,S)}, {\tt (O,S,P)}, and {\tt (O,P,S)}. In a naive
comparison, the VP scheme (sorted by subject) can be seen as an equivalent 
representation to the index {\tt (S,P,O)} in Hexastore. 
%%Hexastore also performs dictionary encoding, so the triples are regarded as 
%%three-ID groups. 
Thus, Hexastore stores triples in a combination of sorted sequences that
requires, in the worst case, 5 times the space used to index the full dataset
in a single triples table. This is because some sequences can be shared between 
different indexes (for instance, the object sequence is interchangeably used
in the indexes {\tt SPO} and {\tt PSO}). The Hexastore organization ensures 
primitive resolution for all triple patterns and also that the first step in 
pairwise joins can be always implemented as fast merge joins. However, its 
large storage requirements slow down Hexastore when representing 
large datasets, because it is implemented as an {\em in-memory} solution.

\paragraph{RDF3X}\cite{NW:10} It goes one step further and 
introduces index compression to reduce the spatial requirements reported 
above. In contrast to Hexastore, RDF3X creates its indexes over a single 
``giant triples table'' (with columns {\tt v$_1$,v$_2$,v$_3$}), and stores 
them in a (compressed) clustered B$^+$-tree. Triples, within each index, are
lexicographically sorted
%%\footnote{RDF3X also performs dictionary encoding, so the ordering is 
%%carried out on the element IDs.} 
allowing SPARQL patterns to be converted into range scans.

The collation order implemented in the RDF3X table causes neighboring triples
to be very similar. In most cases, neighboring triples share the values in
{\tt v$_1$} and {\tt v$_2$}, and the increases in {\tt v$_3$} are very small. 
This fact facilitates differential compression to represent a given
triple with respect to the previous one. This scheme is leaf-oriented within 
the B$^+$-tree, so the compression is individually applied on each leaf. 
Although the authors test some well-known bitwise codes ($\gamma$-codes, 
$\delta$-codes, and Golomb codes \cite{S:07}), they finally apply a bytewise
code specifically designed for diferential triples compression. This technique
ensures highly-efficient decompression with a slight spatial overhead with 
respect to the most effective codes. Finally, it is worth noting that RDF3X
also manages aggregated indexes {\tt (SP)}, {\tt (PS)}, {\tt (SO)}, {\tt (OS)},
{\tt (PO)}, and {\tt (OP)}, which store the number of ocurrences of each pair
in the dataset. RDF3X also contributes with a RISC-style query processor that mainly
relies on merge joins over the sorted indexes. Besides, it implements a query 
optimizer mostly focused on join ordering in its generation of execution plans.

RDF3X reports a very efficient performance that outperforms SW-Store by a 
large margin. These results make it a leading reference in the area. 
However, despite its compression achievements, the spatial requirements in 
RDF3X remain very high. This involves an indirect overhead to the querying 
performance because large amounts of data need to be transferred from 
disk to memory, and this can be a very expensive process with respect to the
query resolution itself \cite{SGKNM:08,SHKLP:08}.

\paragraph{BitMat}\cite{ACZH:10} It follows the idea of managing 
compressed indexes, but it goes another step further and proposes querying 
algorithms that directly perform on the compressed representation. BitMat 
introduces an innovative compressed bit-matrix to represent the RDF structure.
It is conceptually designed as a bit-cube {\tt S$\times$P$\times$O}, but its 
final implementation slices to get two-dimensional matrices: {\tt SO} and 
{\tt OS} for each predicate {\tt P}, {\tt PO} for each subject {\tt S}, and 
{\tt PS} for each object {\tt O}. These matrices are run-length \cite{S:07} 
compressed by taking advantage of their sparseness. Two additional bitarrays 
are used to mark non-empty rows and columns in the bitmats {\tt SO} and 
{\tt OS}. The results reported for BitMat show that it only overcomes the 
state of the art for low selectivity queries. However, it is an interesting 
achievement because it demonstrates that avoiding materializaton of intermediate
results is a very significative optimization for these queries.

\bigskip

Finally, hybrid \cite{SEH:12} and full in-memory stores \cite{JK:05,BGZPS:11} 
%%\cite{JK:05,SK:06,BGZPS:11} 
represent an emerging alternative in this scenario, but their current results 
are limited for managing small datasets, as previously shown for Hexastore.
Their scalability is clearly compromised by the use of structures, like indexes
and hash tables, that demand large amounts of memory. However, some semantic 
applications, such as {\em inference-based} ones, claim for scalable in-memory 
stores because they perform orders of magnitude faster if the entire dataset is
in memory \cite{HAR:11}, and they also support a higher degree of reasoning. 
New opportunities arise for in-memory stores thanks to the advances in 
distributed computing. This class of solutions, recently studied 
\cite{UMB:10,HAR:11} on the MapReduce framework, allows arbitrarily large RDF
data to be handled in main memory because more nodes can be added to a cluster
when more resources were necessary. However, these systems still require 
further research to ensure efficient RDF exchanging \cite{FMPGPA:13,HDT:11}, as
well as efficient performance in each node.

%% file: cds.tex
\section{Succinct Data Structures}
%\section{Background}
\label{s:cds}

Succinct data structures \cite{NM:06} aim at representing data (e.g., sets, 
trees, hash tables, graphs, texts) using as little space as possible.
They are able to approach the information theoretic minimum 
space required to store the original data, but also retain direct access to 
the data. These features yield competitive overall 
performance, because they can implement specific functionality in faster levels
of the memory hierarchy due to the spatial reductions obtained. This section 
covers the basic concepts about the succinct data structures involved in our 
approach.

\subsection{Binary Sequences}
Binary sequences ({\em bitstrings}) are the basis of many
succinct data structures. A bitstring $\mathcal{B}[1,n]$ stores a sequence of
$n$ bits and must provide efficient resolution for three basic operations:

\begin{itemize}
  \item{\tt rank$_a(\mathcal{B},i)$} counts the occurrences of the bit $a$ in 
    $\mathcal{B}[1,i]$.
  \item{\tt select$_a(\mathcal{B},i)$} locates the position for the $i$-th 
     occurence of $a$ in $\mathcal{B}$.
  \item{\tt access$(\mathcal{B},i)$} returns the bit stored in 
     $\mathcal{B}[i]$.
\end{itemize}

All these operations can be resolved in constant time using $n+o(n)$ bits of
total space: $n$ bits for $\mathcal{B}$ itself, and $o(n)$ additional bits for
the structures used to answer the queries. In this paper, we consider an 
implementation \cite{GGMN:05} which uses, in practice, $5\%$ extra space on top
of the original bitstring size and provides fast query resolution.

\subsection{Directly Addressable Codes (DACs)}

The use of variable-length codes is the basic principle of data compression: 
the most frequent symbols are encoded with shorter codewords, whereas longer
codewords are used for representing less frequent symbols. However, 
variable-length codes complicate random access to elements in the 
compressed sequence, which is required in many practical scenarios (as those
studied in this paper) for efficient retrieval. 
Directly Addressable Codes (DACs) \cite{BLN:09b,L:11} are a practical solution 
to this problem.

DACs start from a variable-length encoding of the symbols in a sequence. 
Each codeword (a variable-length bit sequence) is accommodated in a number of 
fixed-length chunks, using as many chunks as necessary, and thus the encoded
sequence can be regarded as a sequence of chunks.
This sequence is rearranged in several levels: the 
first one concatenates the first chunk of all the codewods in the sequence, the 
second level concatenates the second chunk of all codewords of length more than
one, and so on until the chunks of the longest codewords are processed.
Two structures are used for representing the information in each level: an 
array $\mathcal{A}$ concatenates the chunks corresponding to this level, and a
bistring $\mathcal{B}$, which stores one bit for each element in $\mathcal{A}$,
indicates whether each chunk is the last within its codeword.

\begin{figure}
  \centering
  \hspace{0.25cm}
  \includegraphics*[scale=0.35]{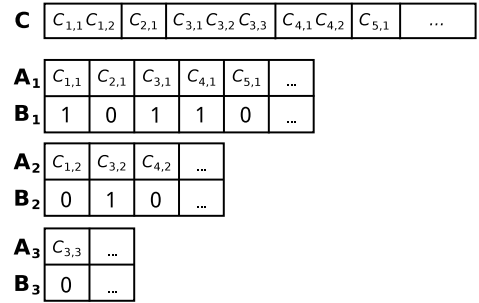}
  \caption{\label{fig:dac} Example of the DAC-based representation of the
    sequence $\mathcal{C}$.}
\end{figure}

\begin{example}
Figure \ref{fig:dac} illustrates this reorganization. The compressed sequence
$\mathcal{C}$ comprises five symbols: the first one is encoded with a 
2-chunk codeword ($\mathcal{C}_{1,1}$, $\mathcal{C}_{1,2}$), the second one 
with a 1-chunk codeword ($\mathcal{C}_{2,1}$), and so on. The first DAC level 
stores in $\mathcal{A}_1$ all the first chunks of the codewords
in $\mathcal{C}$: $[\mathcal{C}_{1,1},\mathcal{C}_{2,1}, \mathcal{C}_{3,1},
\mathcal{C}_{4,1},\mathcal{C}_{5,1}]$, whereas the bitstring $\mathcal{B}_1$ 
list all corresponding bits: $[10110]$. As can be seen, 0-bits indicate that the
corresponding codewords finish in the current level (the second and the 
fifth codewords are fully represented with a single chunk), whereas 1-bits mean
that the codewords continue in the next level (the first, third, and 
fourth codewords are continued in the second level). Thus, the second level
represents the second chunks of these codewords: 
$\mathcal{A}_2=[\mathcal{C}_{1,2},\mathcal{C}_{3,2}, \mathcal{C}_{4,2}]$, and
gives the information about which finish: $[010]$. Finally, the third 
level encodes the last chunk of the third codeword.
\end{example}

DACs enable direct access to any element in the encoded sequence.
To access the codeword at position $i=i_1$, 
$\mathcal{B}_1[i_1]$ is firstly checked. If $\mathcal{B}_1[i_1]=0$, this is 
the last chunk and the value is fully represented in the current level, so
$\mathcal{C}[i]=\mathcal{A}_1[i_1]$. Otherwise ($\mathcal{B}_1[i_1]=1$), the
following chunks must be fetched. The codeword is continued in the position
$i_2=rank_1(\mathcal{B},i_1)$ of the second level. As before, the bitstring 
is firstly checked: if $\mathcal{B}_2[i_2]=0$, this is the last chunk. In 
this case, the codeword value is obtained as 
$\mathcal{C}[i]=\mathcal{A}_1[i_1]+\mathcal{A}_2[i_2]·2^b$, where $b$ is the
chunk length (in bits). If $\mathcal{B}_2[i_2]=1$, the process continues 
iteratively until the codeword is fully extracted.

Accessing a codeword in a DAC compressed sequence takes $O(\log(M)/b)$ time in 
the worst case, where $M$ is the longest codeword length. However, this access 
time is lower for elements with shorter codewords, and these are the most 
frequent ones.

\subsection{K$^2$-trees}

The k$^2$-tree \cite{BLN:09,L:11} is a succinct data structure for
graph representation. It models a graph of $n$ vertices through its (binary) 
adjacency matrix, $\mathcal{M}$, of size $n \times n$. Thus, 
$\mathcal{M}[i,j]=1$ iff the vertices represented in the $i$-th 
row and the $j$-th column are related.

The k$^2$-tree leverages sparseness and clustering features, which arise in some
classes of real-world graphs (such as Web graphs \cite{BLN:09} and social 
networks \cite{CL:11}), to achieve compression. These features imply the 
existence of 
large ``empty'' areas (all cells have value {\tt 0}), and the k$^2$-tree 
excels at compressing
them. Conceptually, the k$^2$-tree subdivides\footnote{The division strategy 
is similar to that proposed in the MX-Quadtree 
\cite[Section 1.4.2.1]{S:06}.} the matrix into $k^2$ sub-matrices of the same 
size, which results in $k$ rows and $k$ columns of sub-matrices of size $n^2/k^2$. 
Each of those $k^2$ sub-matrices is represented in the tree using a single bit 
that is appended as a child of the root: a bit {\bf 1} is used for 
representing those sub-matrices containing at least one cell with value $1$, 
whereas a {\bf 0}-bit means that all cells in the corresponding sub-matrix are 
$0$. Once this first level is built, the method proceeds recursively for each 
child with value $1$ until sub-matrices full of 0s or the last level of the tree
are reached. This process results in a non-balanced k$^2$-ary tree in which the 
bits in its last level correspond to the cell values in the original matrix. 
If the number of rows and columns in the
adjacency matrix is not a power of $k$, the matrix is expanded to the right 
and bottom with 0s, obtaining an extended matrix of size 
$n' \times n', n'=k^{\lceil \log_k n \rceil}$. This expansion causes just a 
little overhead because of the $k^2$-tree ability to compress the large 
areas of 0s created after the expansion.

\begin{figure}
\centering
   \includegraphics*[bb=0 365 1200 840,scale=0.295]{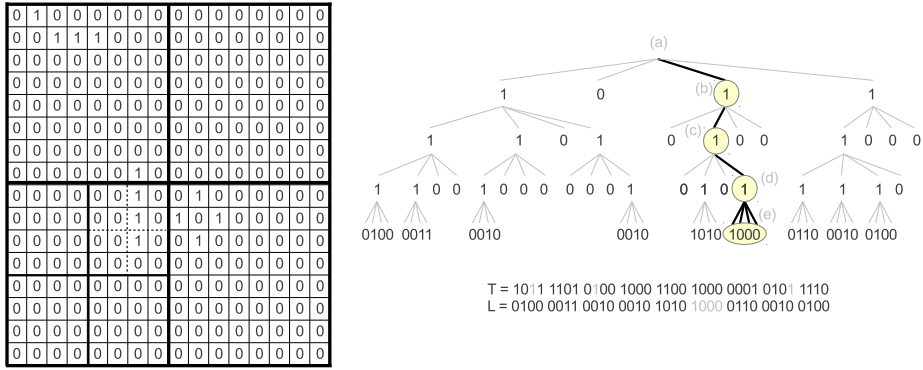}
  \caption{\label{fig:k2tree} Example of k$^2$-tree for an adjacency matrix of
    size $16 \times 16$.}
\end{figure}

The k$^2$-tree is implemented in a very compact way using two bitstrings: 
{\bf T} (tree) and {\bf L} (leaves). {\bf T} stores all the bits in 
the k$^2$-tree except those stored in the last level. The bits are placed following
a levelwise traversal: first the k$^2$ binary values of the children of the 
root node, then the values of the second level, and so on. This configuration 
enables the k$^2$-tree to be traversed by performing efficient {\tt rank} and
{\tt select}
operations on the bitstring. On the other hand, {\bf L} stores the last level of the 
tree, comprising the cell values in the original matrix. Although {\bf L} was 
originally devised to be implemented with a bistring, a recent 
improvement stops the decomposition when the matrices reach size
$k_L \times k_L$ and uses DACs to compress them according to frequency while
retaining fast direct access to any leaf sub-matrix \cite{L:11}. 

Besides of its compression ability, the k$^2$-tree provides various
navigational operations on the graph. In particular, for a given vertex $v$, 
the k$^2$-tree supports the operations of retrieving $(i)$ all the vertices 
pointed by $v$ ({\em direct neighbors}), and $(ii)$ all the vertices that
point to $v$ ({\em inverse neighbors}). 
Additionally, {\em range queries} (retrieving all the connections within a 
sub-matrix), and the fast {\em check of a given cell value} are also supported
by the k$^2$-tree.

Conceptually, direct neighbors retrieval, for a vertex $v_i$, 
requires finding all the cells with value $1$ in the $i$-th row. Symmetrically,
the inverse neighbors of $v_i$ are retrieved by locating all the 1s in the 
$i$-th column. Both operations are efficiently implemented on a top-down 
traversal of the tree, requiring O(1) time per node visited (and $O(n)$ overall
time in the worst-case, but possibly less in practice). This traversal
starts at the root of the k$^2$-tree, $pos=0$, and visits in each step the
children of all nodes with value 1 in the previous level and whose matrix area
is not disjoint from the area one wants to retrieve. Given a node,
represented at the position $pos_i$ of {\bf T}, its $k^2$ children 
are represented consecutively from the position 
$pos_i=rank_1(T,pos) \cdot k^2$ of {\bf T}:{\bf L}.

\begin{example}
Figure \ref{fig:k2tree} shows a $16\times 16$ adjacency matrix
(left) and the k$^2$-tree (right) representing it, using $k=2$.
The configurations for the the two bitstrings, {\bf T} and
{\bf L}, implementing the k$^2$-tree, are also shown at the bottom of the figure.
The matrix is conceptually 
divided into $2^2=4$ sub-matrices. In the first step, the sub-matrices are of
size $8 \times 8$. Assume we are interested in retrieving the 
{\em direct neighbors} of the $11$-th vertex, so we need to find all
the cells with value $1$ in the $11$-th row of the adjacency matrix. The 
first step starts at the root of the k$^2$-tree, $pos=0$, and computes the
children overlapping the eleventh row. These are the third and the fourth 
children (representing the sub-matrices at the bottom of the 
original adjacency matrix), and these are respectively represented in 
$\mathbf{T}[2]$ and $\mathbf{T}[3]$ (assuming that positions in {\bf T} are numbered from 0). In both
cases, $\mathbf{T}[2]$ and $\mathbf{T}[3]$ have value 1, so both children must be traversed.
For simplicity we only detail the procedure for the {\em third} child, so 
now $pos=2$. The second step first computes the position representing the
first child of the current vertex: 
$pos=rank_1(\mathbf{T},2) \cdot 2^2=2 \cdot 4=8$, and checks the value of the 
$k^2=4$ bits stored from $\mathbf{T}[8]$: $[0100]$. In this case, only the second child
(represented at $pos=9$) has value 1, so this is the node to be visited in 
the third step. The children for this node are located from 
$pos=rank_1(\mathbf{T},9) \cdot 2^2=7 \cdot 4=28$, and contain values $[0101]$. 
Although the second child is 1, this is not a valid match for our query because
it has no intersection with the $11$-th row. This means that only 
the fourth child (represented at $pos=31$) is visited in the fourth step. 
The new position $pos=rank_1(\mathbf{T},31) \cdot 2^2=14 \cdot 4=56$ is larger 
than $\vert \mathbf{T} \vert = 36$, so it represents a leaf. Thus, the $k^2$ resulting
leaves must be checked from position $56-36=20$ of {\bf L}. The bits $[1000]$ 
represent this submatrix, so one connection is found for the $11$-th row, and it is
the $7$-th column.
\end{example}

Similar algorithms implement the extended
functionality. Checking the value of a given cell also uses a 
recursive descent, but it only visits the single appropiate child, in each level, for
the given query. Range queries are performed similarly, but each step 
visits all the children representing the rows and columns involved in the query. More 
details about both algorithms can be found in \cite{L:11}.

%% file: k2triples.tex
\section{Full-In-Memory Vertical Partitioning on k$^2$-{\sc triples}}
\label{s:k2t1}

This section describes how the k$^2$-tree structure can be applied to the problem 
of RDF storage. Our approach is called {k$^2$-{\sc triples}}. We firstly 
perform a
specific {\em dictionary encoding} that allows triples to be managed as 
three-ID groups: $(id_1, id_2, id_3)$, in which $id_1$ is the integer value 
that identifies the subject in the dictionary, $id_2$ identifies the 
predicate, and finally $id_3$ identifies the object. This decision simplifies
{\em data partitioning} on k$^2$-trees because a direct correspondence can be
established between rows and columns in the adjacency matrix and subject and 
object IDs in the dictionary.

\subsection{Dictionary Encoding}

Dictionary encoding is a common preliminary step performed before 
data partitioning. All different terms used in the dataset are firslty 
extracted from the dataset and mapped to integer values through a dictionary
function. As explained above, this allows long terms occurring in the RDF triples
to be replaced by short IDs referencing them within the dictionary. This simple
decision greatly compacts the dataset representation, and mitigates scalability
issues.

We propose a dictionary organization comprising four independent categories, in
which terms are usually organized in lexicographic order 
\cite{ACZH:10,MPFC:12,FMPGPA:13}:

\begin{itemize}
  \item{\bf Common subjects and objects} ({\tt SO}) organizes all the terms that
    play both subject and object roles in the dataset. They are mapped to the
    range {\tt [1, $\vert$SO$\vert$]}.
  \item{\bf Subjects} ({\tt S}) organizes all the subjects that do not play an 
    object role. They are mapped to 
    {\tt [$\vert$SO$\vert$+1, $\vert$SO$\vert$+$\vert$S$\vert$]}.
  \item{\bf Objects} ({\tt O}) organizes all the objects that do not play a 
    subject role. They are mapped to 
    {\tt [$\vert$SO$\vert$+1, $\vert$SO$\vert$+$\vert$O$\vert$]}. Note this
interval overlaps with that for {\tt S}, since confusion cannot arise.
  \item{\bf Predicates} ({\tt P}) maps all the predicates to {\tt [1,$\vert$P$\vert$]}.
\end{itemize}

In this way, terms playing subject and object roles are represented only once. 
Moreover, the intervals for subjects and objects are contiguous.
This decision, on the one hand, provides a dictionary size reduction as it
prevents the duplicate representation of these terms. This is a significant 
reduction if we consider that up to $60\%$ of the terms in the dictionary are
in the {\tt SO} area for real-world datasets \cite{MPFC:12}. On the other 
hand, this four-category organization improves performance 
for subject-object joins because all their possible matches are 
elements playing both subject and object roles, and all of them are in the range 
{\tt [1, $\vert$SO$\vert$]}. Thus, this join resolution is concentrated on the
area of {\tt SO} and avoids querying for the remaining subjects and objects.

How the dictionary is finally implemented is orthogonal to the problem
addressed in this paper, and any existing technique within the state of the art
could be adapted for managing our organization. Nevertheless, we emphasize that
RDF dictionaries take up to 3 times more space than that required for 
representing the triples structure \cite{MPFC:12}, so compressed dictionary 
indexes are highly desirable for efficient and scalable management of huge RDF
datasets.

\begin{figure}
  \centering
  \includegraphics*[bb=20 299 575 564, scale=0.5]{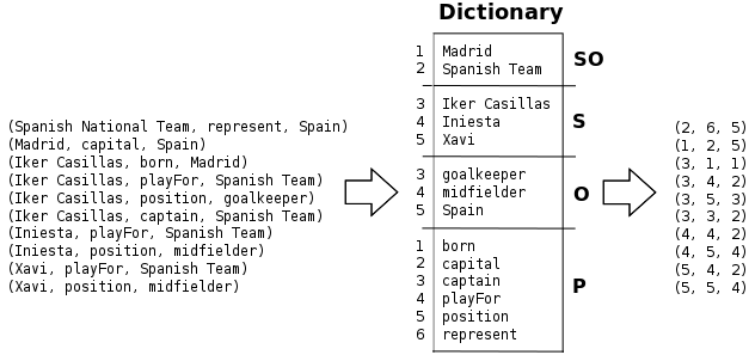}
  \caption{\label{fig:dict} Example of {\em dictionary encoding} on k$^2$-{\sc triples}.}
\end{figure}

\begin{example}
Figure \ref{fig:dict} illustrates our dictionary organization over the RDF 
excerpt used in Figure \ref{fig:exrdf}. As can be seen, the terms {\tt Madrid} 
and {\tt Spanish Team} (playing as subject and object) are respectively 
identified with the values 1 and 2, the three subjects are represented in the 
range {\tt [3,5]}, and equally the three objects are identified with the same 
values: {\tt \{3,4,5\}}. Finally, the six predicates used in the example are 
identified in the range {\tt [1,6]}. On the right of the figure,
the ID-based representation of the original triples is shown.
\end{example}

%%%%%%%%%%%%%%%%%%%%
\begin{figure}
  \centering
  \includegraphics*[bb=0 340 565 525,scale=0.6]{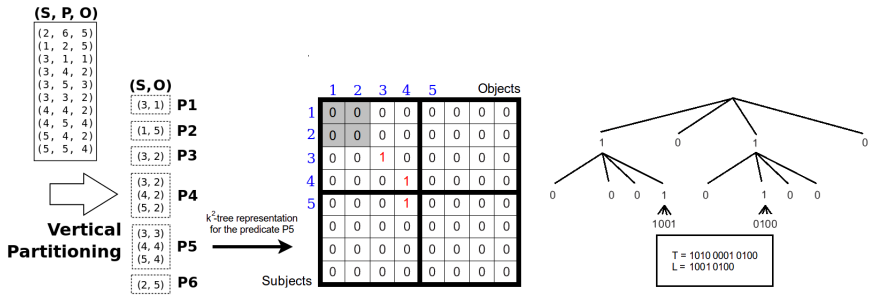} %18 393 505 556
  \caption{\label{fig:vert} Vertical Partitioning on k$^2$-{\sc triples} (the 
    parameter $k$ is set to 2).}
\end{figure}

\subsection{Data Partitioning}

k$^2$-{\sc triples} models RDF data following the well-known {\em vertical
partitioning} approach. This scheme reorganizes a dataset into 
$\vert${\tt P}$\vert$ disjoint subsets that contain all the triples related to a
given predicate. Thus, all triples in a subset can be rewritten as pairs of 
subject and object (S,O), because the corresponding predicate is implicitly 
associated to the given subset. 

Each subset is independently indexed in a single k$^2$-tree that represents
subjects and objects as rows and columns of the underlying matrix. That is, 
each k$^2$-tree models an adjacency matrix of 
$\vert${\tt SO}$\vert + \vert${\tt S}$\vert$ rows and 
$\vert${\tt SO}$\vert + \vert${\tt O}$\vert$ columns. In practice, as commented
above, the k$^2$-tree is extended to the next power of $k$ to obtain a square
matrix of size $n' \times n', n'= k^{\lceil \log_k(|SO|+\max(|S|,|O|)) \rceil}$.

Finally, it is worth noting that all the k$^2$-trees used in our approach are 
physically built with a hybrid policy that uses value $k=4$ up to the level $5$
of the tree, and then $k=2$ for the rest \cite{BLN:09,L:11}. The leaves, regarded as submatrices of size $8 \times 8$, are encoded using DACs \cite{BLN:09b,L:11}.

\begin{example}
Figure \ref{fig:vert} (left) 
shows the vertical partitioning of our excerpt of RDF. As can be seen, six 
different subsets are obtained (one for each different predicate), and the triples
are rewritten as pairs (S,O) within them. For example, the triples for
the predicate {\bf 5}: {\tt (3,5,3), (4,5,4), (5,5,4)}, are rewritten as pairs:
{\tt (3,3), (4,4), (5,4)}, and they are then managed within the subset {\bf P5}
associated to the predicate 5.

The right side of the figure shows the adjacency matrix underlying 
to the k$^2$-tree used for representing the subset P5. The shadow cells 
are the area {\tt SO} in which elements playing as subjects and nodes are
represented. Note that only the first five rows (for the five existing 
subjects) and the five first columns (for the five existing objects) are really
used, hence all the triples are stored in these ranges. In this case, 1-bits are 
found in the cells {\tt (3,3), (4,4), (5,4)} in which the triples {\tt (3,5,3),
(4,5,4), (5,5,4)} are represented. As can be seen, the resulting matrix 
contains a very sparse distribution of 1-bits, and this is the best scenario 
for k$^2$-trees because of their ability to compress large empty areas.
\end{example}

\subsection{Indexing Predicates for Subjects (SP) and Objects (OP)}

The resolution of queries involving variable predicates is the main 
weakness of systems that implement vertical partitioning.
In our case, all k$^2$-trees must be 
traversed for resolving triple patterns with unbounded predicate (see next
section). This is hardly scalable when a large number of predicates is used in
the dataset. In this section, we enhance k$^2$-{\sc Triples} in order to 
minimize the number of k$^2$-trees that must be traversed for resolving triple
patterns involving variable predicates. 

The triple pattern classification, given in Section \ref{s:state}, shows that
{\tt (?S,?P,?O)} is the only pattern with unbounded predicate that provides 
neither value for the subject nor the object. However, this pattern always 
scans the full dataset to retrieve all the triples within it, and no specific
optimizations are possible for it. Thus, we can leverage that subject and/or
object values are provided, and use them to optimize the number of k$^2$-trees
that must be traversed for resolving the remaining patterns with unbounded 
predicate. This is achieved through two specific indexes:

\begin{itemize}
  \item The {\em Subject-Predicate} (SP) index organizes the lists of 
    all the different predicates related to each subject in the dataset.
  \item The {\em Object-Predicate} (OP) index organizes the lists of 
    all the different predicates related to each object in the dataset.
\end{itemize}

Empirical results \cite{FMPGPA:13} show that the average size of 
these lists of predicates for subjects and objects is, at most, one order of 
magnitude less than the number of total predicates used in real-world datasets.
This fact not only ensures a great improvement for queries with unbounded 
predicate, but also implies a limited additional space for SP and OP indexes. 
Thus, this enhanced revision of k$^2$-{\sc Triples} also retains the original 
aim of obtaining a compact RDF representation.

\begin{figure}
  \centering
  \includegraphics*[bb=18 313 577 551,scale=0.6]{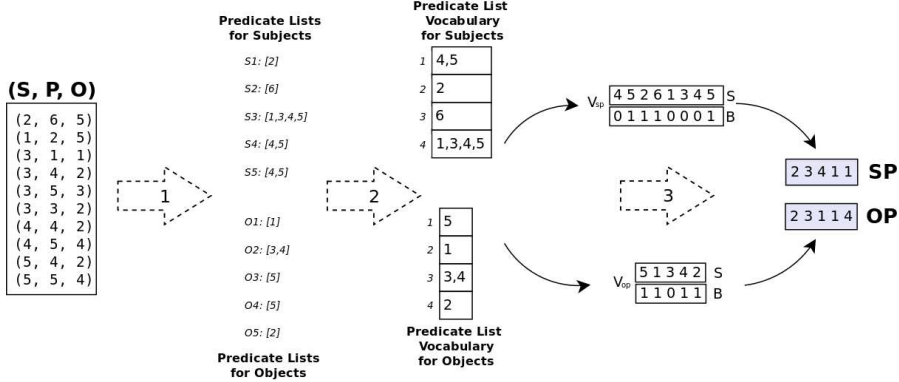}
  \caption{\label{fig:enh} Building SP and OP indexes.}
\end{figure}

Both SP and OP indexes rely on a compact representation of their 
{\em predicate lists}. The predicate list, for a given subject, organizes
all the different predicates related to it. These predicate lists are common for
some subjects, and this can be leveraged to achieve space savings. For this
purpose, we obtain the set of all the different predicate lists (we referred this
set to as the {\em predicate list vocabulary}), and sort them according to their 
frequency. In this way, the predicate lists appearing in more subjects are represented in 
the first positions of the vocabulary and are encoded with smaller codewords.
A similar reasoning applies for objects. We respectively refer to 
{\bf V$_{sp}$} and {\bf V$_{op}$} as the predicate list vocabularies for subjects
and objects.

\begin{example}
The arrow 1 in Figure \ref{fig:enh} shows the predicate lists 
obtained for the subjects and objects in the RDF excerpt used in 
the previous examples. As can be seen, the subject {\tt 1} is related to a 
1-element list containing the predicate {\tt 2}; the list for the subject 
{\tt 2} contains the element {\tt 6}; for the subject {\tt 3}, its predicate
list contains four elements: {\tt 1,3,4,5}; and the subjects {\tt 4} and 
{\tt 5} are related to 2-element predicate lists containing the elements 
{\tt 4,5}. The arrow 2 illustrates how predicate lists vocabularies are 
obtained. Let us consider the case of subjects: the list {\tt 4,5} is 
represented in the first position because it is related to two different
subjects ({\tt 4} and {\tt 5}), whereas the other lists are only related to a
single subject. The case of objects is similar: the list {\tt 5}
is related to the objects {\tt 3} and {\tt 4}, and the remaining lists appear
only once.
\end{example}

We propose a succinct vocabulary representation based on the following two
structures:

\begin{itemize}
  \item An integer sequence $\mathcal{S}$ that concatenates all the predicate 
    lists according to their frequency. Thus, the most frequent lists appear
    at the beginning of the sequence, whereas the least frequent ones are at the
    end. Each element in $\mathcal{S}$ takes $\log (\vert \mathtt{P} \vert)$ bits of
    space.
  \item A bitstring $\mathcal{B}$ that delimits and identifies predicate 
    lists within the vocabulary. That is, the $i$-th 1-bit in the position $p$ 
    of the bit sequence means that the predicate list identified as $i$ finishes 
    in the $p$-th position of $\mathcal{S}$. 
\end{itemize}

$\mathcal{B}$ enables efficient list extraction within the vocabulary (let us 
assume that $\mathtt{select}_1(\mathcal{B}, 0)=1$), because of the $p$-th
predicate list is stored in $\mathcal{S}[i,j]$, where
$i=\mathtt{select}_1(\mathcal{B},p-1)+1$, and
$j=\mathtt{select}_1(\mathcal{B},p)-1$.

This representation allows the SP and OP indexes to be easily modelled as
integer sequences. We detail SP, but the same representation applies to OP.
The index SP is modelled as a sequence of integer IDs of length 
$\vert \mathcal{S} \vert$. In this way, the $p$-th value in SP (referred to as 
$\mathrm{SP}[p]$) contains the ID of the predicate list related to the subject
$p$, and it can be extracted from V$_{sp}$ by using the simple 
{\em extraction} process explained above. The elements of index SP are 
finally represented 
using DACs. This retains direct access to any position in the index and
also leverages the frequency distribution of predicate lists to achieve 
compression. Note that DACs assign shorter codewords to smaller integers and 
these are used for representing the most frequent lists within the vocabulary.

\begin{example}
The arrow 3, in Figure \ref{fig:enh}, illustrates the final SP and OP index
configurations. As can be seen, SP lists the IDs {\tt [2,3,4,1,1]}. This means that 
the first subject is related to the second predicate list, the second subject 
to the third list, and so on. For instance, if we want to extract the list of
predicates related to the subject {\tt 3}, we firstly retrieve its ID as 
$\mathrm{SP}[3]=4$. Thus, the fourth list must be extracted from V$_{sp}$. 
This is represented in $\mathcal{S}$ from the position 
$i=\mathtt{select}_1(\mathcal{B},3)+1=4+1=5$ to the position
$j=\mathtt{select}_1(\mathcal{B},4)=8$, and contains the predicates 
{\tt 1,3,4}, and {\tt 5}.
\end{example}

\section{Triple Pattern Resolution over k$^2$-{\sc triples}}
\label{s:k2t2}

Triple patterns are the basic lookup unit on RDF triples; more complex 
SPARQL queries can be translated into plans involving triple pattern 
resolution. Thus, RDF retrieval strongly relies on the performance 
achieved for triple pattern resolution. This is one of the main strengths of our
approach, because k$^2$-{\sc triples} can answer all patterns on the 
highly-optimized operations provided by the k$^2$-tree structure:

\begin{itemize}
  \item{\tt (S,P,O)} is directly implemented on the operation that
    {\em checks the value of a given cell} in the k$^2$-tree. That is, 
    the triple {\tt (S,P,O)} is in the dataset iff the cell {\tt (S,O)} (in
    the matrix representing the subset of triples associated to the predicate 
    {\tt P}) contains the bit $1$. 
    This operation returns a boolean value and it is usually required within
    {\tt ASK} queries.
  \item{\tt (S,?P,O)} generalizes the previous pattern by checking the value
    of the cell {\tt (S,O)} in all the k$^2$-trees. The result is an ID-sorted 
    list of all the predicates whose k$^2$-tree contains a 1 in this cell.
    The process can be sped up by first intersecting the predicate lists
    of SP and OP respectively associated to {\tt S} and {\tt O}, obtaining a list of 
    predicates $\mathtt{P}_i$ that contain objects related to {\tt S} as well
    as subjects related to {\tt O}. Then, only the k$^2$-trees of those 
    $\mathtt{P}_i$ need be considered for pairs {\tt (S,O)}. 
  \item{\tt (S,P,?O)} can be seen as a forward navigation from {\tt S} to all
    the objects related to it through predicate {\tt P}. Thus, it is equivalent to a 
    {\em direct neighbors} retrieval that locates all the columns with value 1 in
    the row associated to the subject {\tt S} within the k$^2$-tree for 
    {\tt P}. The objects matching the pattern are returned in sorted order.
  \item{\tt (S,?P,?O)} generalizes the previous pattern by performing direct
    neighbor retrieval in all the k$^2$-trees. In this case, the result comprises
    many ID-sorted lists of objects for the predicates related to {\tt S}.
    This is sped up by using the information stored in the SP index.
    A subject-based query on this SP index returns the predicate
    list containing all predicates {\tt P$_i$} related to the subject {\tt S}. 
    Thus, direct neighbors retrieval is only performed on the 
    {\tt $\vert$P$_i \vert$} k$^2$-trees modeling the predicates within the list.
  \item{\tt (?S,P,O)} corresponds to a backwards navigation from {\tt O} to all
    the subjects related to it through {\tt P}. This is equivalent to a 
    {\em reverse neighbors} retrieval that locates all the rows with value 1 in 
    the columns associated to the object {\tt O} within the k$^2$-tree for 
    {\tt P}. The subjects matching the pattern are returned in sorted order.
  \item{\tt (?S,?P,O)} generalizes the previous pattern by performing reverse
    neighbors retrieval in all the k$^2$-trees. In this case, the result comprises
    many ID-sorted lists of subjects for the predicates related to {\tt O}.
    An object-based query on the OP index speeds up the query by restricting the
    predicate list to those {\tt P$_j$} with which object 
    {\tt O} relates to some subject. 
  \item{\tt (?S,P,?O)} is equivalent to retrieving all the values 1 in the 
    k$^2$-tree associated to the predicate {\tt P}. This is easily implemented
    by a range query performing a full top-down traversal that retrieves all 
    the pairs {\tt (S,O)} in the structure.
  \item{\tt (?S,?P,?O)} generalizes the previous pattern by obtaining all the 1s
    in all the k$^2$-trees used for representing the predicates in the 
    dataset.
\end{itemize}

\section{Join Resolution over k$^2$-{\sc triples}}
\label{s:k2t3}

The core of SPARQL relies on the concept of {\em Basic Graph Pattern} (BGP), 
and its semantics to build conjunctive expressions by {\em joining} triple 
patterns through shared variables. BGPs are reordered and partitioned into 
pairs of triple patterns sharing exactly one variable. Thus, the performance of
BGPs is clearly determined by the algorithms available for joining these
patterns, and also for the optimization strategies to order the joins. This 
second topic is orthogonal to this paper and is not addresed. 

%% Traditional optimizers for disk-based representations mainly focus on 
%% reducing I/O costs because these are the slowest operations in semantic 
%% queries \cite{G:11}. Thus, in the current scenario, specific optimization
%% techniques must be designed to address the different situation arisen for
%% in-memory RDF representations. Optimizers based on selectivity estimation 
%% \cite{SSBKR:08} seem the best positioned.

k$^2$-{\sc triples} provides {\em subject-subject} and 
{\em object-object} join resolution. This overcomes traditional vertical 
partitioning, which only gives direct support for subject-subject joins, and 
requires an additional object index for efficient resolution of object-object
joins. Besides, k$^2$-{\sc triples} also supports {\em subject-object} joins. 
These are efficiently implemented by considering only the common area {\tt SO} in
which nodes playing as subjects and objects are exclusively represented. Our 
native support for cross-joins is a significant improvement with respect to
traditional vertical partitioning, in which framework cross-joins are 
described as rather expensive and
inefficient operations \cite{AMMH:07}. This fact is a clear weakness for these
traditional solutions because cross-joins are the basis for implementing the 
common {\em path expressions} queries. In \cite{AMMH:07}, the path expression
query problem is tackled in a non-general way: the results of only some 
selected paths are  precalculated and stored for their efficient querying.
Finally, it is worth noting that operations involving predicates as join 
variables are underused in practical terms \cite{AFMP:11b}. 
%% Although they are
%% not addressed in this paper, {\em predicate-predicate} joins can be directly 
%% resolved on the k$^2$-{\sc triples} configuration, and cross joins 
%% structure mapping predicate IDs and their respective counterparts as subjects
%% and/or objects.

This section describes the algorithms and mechanisms implemented on 
k$^2$-{\sc triples} for join resolution. We firstly classify join operations 
and then detail the join algorithms implemented by our approach. The section
ends with the description of the specific mechanisms provided by 
k$^2$-{\sc triples} for resolving each kind of join operation according to 
our different join algorithms.

\subsection{Classifying Join Operations}
\label{s:k2t4}

\begin{figure}
  \centering
  \includegraphics*[bb=20 339 574 528,scale=0.625]{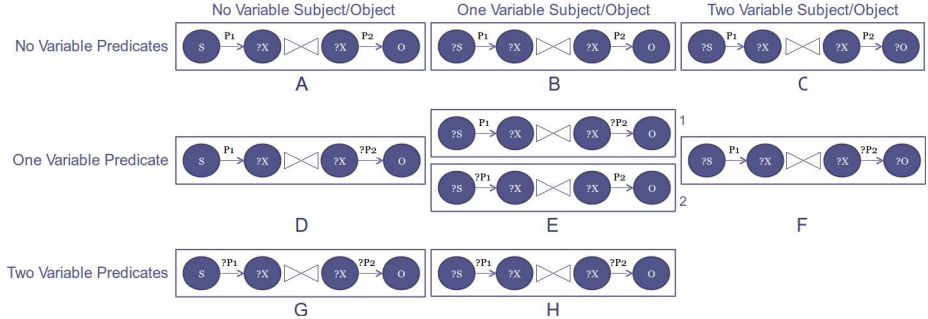}
  \caption{\label{fig:joins} Classification of subject-object joins supported in k$^2$-triples.}
\end{figure}

Figure \ref{fig:joins} classifies the join operations according to the classes
studied in this section. Although all of them refer to subject-object joins,
subject-subject and object-object ones are similarly classified and solved
on the same guidelines. We refer to {\tt ?X} as the join variable in each class.

Join operations are organized, {\em by rows}, according to the state of the
predicates in the two patterns involved in the join:

\begin{itemize}
   \item Row {\em no variable predicates} lists the joins in which both triple 
     patterns provide their predicates (classes {\bf A, B} and {\bf C}).
   \item Row {\em one variable predicate} lists the joins in which one triple 
     pattern provides its predicate, whereas the other one leaves it variable 
     (classes {\bf D, E} and {\bf F}).
   \item Row {\em two variable predicates} lists the joins in which both 
     triple patterns leave as variables their corresponding predicates (classes {\bf G} and {\bf H}).
\end{itemize}

The {\em column}-based classification lists join operations according to the
state of the nodes in the triple patterns. If we consider that the join 
variable is represented in two of these nodes, the remaining two determine the
classes: 

\begin{itemize}
   \item Column {\em no variable subject/object} lists the joins in which
     the value of the two not joined nodes are provided (classes {\bf A, D} and {\bf G}).
   \item Column {\em one variable subject/object} lists the joins in which
     one triple pattern provides its not joined node, whereas the other one leaves 
     it variable (classes {\bf B, E} and {\bf H}). From this perspective, the 
     class E is split into two different subclasses: in {\bf E.1}, one pattern 
     provides its predicate but leaves variable its node, whereas the other 
     pattern provides the node but leaves as variable its predicate; in 
     {\bf E.2}, one pattern is full-of-variables (it does not provide neither
     the node nor the predicate), whereas the other one provides both the node 
     and the predicate.
   \item Column {\em two variable subject/object} lists the joins in which 
     both triple patterns leave as variables their not joined nodes (classes {\bf C} and {\bf F}).
\end{itemize}

It is woth noting that the eventual class {\bf I} is not studied because joins
full-of-variables: {\tt (?S,?P$_1$,?X) (?X,?P$_2$,?O)}, are not used in 
practice.

\subsection{Join Algorithms}
Join algorithms have been widely studied for relational databases \cite{RG:00},
and have been recently reviewed from the perspective of semantic Web databases
\cite{G:11}. We gather this experience to propose three join algorithms 
optimized for performing over k$^2$-{\sc triples}. We will use a simple 
notation where $T_l$ and $T_r$ refer to the left and right triple patterns,
respectively, involved in the join.

%%  \item{\bf Nested-loop join} is the simplest algorithm and its high complexity
%%    restricts its application to small sets of unsorted data. It contains a
%%    nested loop retrieving all solutions $s_l$ for $T_l$, and for each one,
%%    all solutions $s_r$ for $T_r$ are checked. If $s_l$ and $s_r$ can be 
%%    joined, its result is returned.
%%  \item{\bf Hash join} is the fastest general-purpose algorithm whenever the
%%    intermediate solutions are neither sorted nor an already existing index can
%%    be used. It firstly resolves each triple pattern and then hashes their 
%%    intermediate solutions using the same hash function. This way, solutions 
%%    sharing the same hash are stored in the same partition, and equivalent 
%%    partitions for each pattern are finally joined. In practical terms, hash 
%%    join requires enough memory to hold the hash tables.

\paragraph{Chain evaluation} This algorithm relies on the foundations of 
the traditional {\em index join}: it firstly retrieves all the solutions
for $T_l$, and then each one is used for obtaining all the solutions for $T_r$.
%% through an indexed query. Chain evaluation is typically used whenever an 
%% index is provided for the efficient querying of $T_r$.
%%
Our implementation firstly resolves the less expensive pattern (assume it
is $T_l$), and gathers all the values {\tt X$_i$} obtained for
the join variable {\tt ?X}. All these values are then used for replacement in
$T_r$. Note that some of these values can be duplicated and these
must be identified before the replacement. These duplicates may belong to the 
result of range queries or multiple direct/reverse neighbors. 
We implement an adaptive sort \cite{K:73} 
algorithm that merges the results obtained for each predicate leveraging that
these are returned in sorted order.

%% \begin{description}
%%   \item{\em Direct/reverse neighbor queries} can output duplicates if they
%%     are performed within a triple pattern with unbounded predicate: 
%%     {\tt (S,?P$_1$,?X)} or {\tt (?X,?P$_2$,O)}. In both cases, the values for
%%     {\tt ?X} are retrieved in sorted fashion for each predicate, so the final
%%     result set contains ordered subsets ({\bf runs}) of objects (for 
%%     {\tt (S,?P$_1$,?X)}) or subjects (for {\tt (?S,?P$_2$,O)}). An adaptive 
%%     sort \cite{K:73} based algorithm arises as the best choice for detection of
%%     duplicates in this scenario. It leverages the internal order existing in 
%%     the runs to perform in proportional time to the number of these runs rather
%%     than the total number of results retrieved for {\tt ?X}. Adaptive sorting 
%%     is an efficient choice in a practical RDF scenario due to the average 
%%     number of predicates related to a given subject or object remains very low
%%     \cite{FMPG:10}. This means that the number of runs obtained for the cited 
%%     patterns is also low and the operation will be performed in competitive 
%%     time.
%%  \item{\em Range queries}, in patterns {\tt (?S,P,?X)} or {\tt (?X,P,?O)},
%%     output duplicates because an object can be related to many subjects and vice
%%     versa. In these cases, no assumptions can be made about ordering and a 
%%     general quick sort algorithm is used in order to detect duplicates.
%% \end{description}

\paragraph{Independent evaluation} This algorithm implements 
the well-known {\em merge join}: it firstly resolves both triple patterns and 
then intersects their respective solutions, which come sorted by the join
attribute.\footnote{This is done by traversing the k$^2$-tree in the proper
order or by sorting the results afterwards.}
%% This independent evaluation algorithm
%% requires that the solutions for $T_l$ and $T_r$ to be sorted by the join 
%% variable, and the final result set contains the intersection of both groups
%% of solutions. 
%% k$^2$-{\sc triples} implements merge join ``as is'', so it resolves each triple pattern and then merge their solutions.

\paragraph{Interactive evaluation} This algorithm is strongly inspired on 
the Sideways Information Passing (SIP) mechanism proposed by Neumann and Weikum
\cite{NW:09}. SIP passes information on-the-fly between the operands involved 
in the query in a way that the processing performed in one of them can 
feed back the other and vice versa. Thus, both triple patterns within the join
are interactively evaluated and resolved without materialization of 
intermediate results. This interactive evaluation is easily implemented in 
k$^2$-{\sc triples} by means of a coordinated step-by-step traversal performed
on those k$^2$-trees involved in the resolution of each pattern within the 
join. In the next example only two k$^2$-trees are involved, the join attribute
is the subject in both trees, and the predicates and objects are fixed, but all
the other combinations can be handled similarly.

%% \begin{algorithm}[ht!]
%%   \begin{algorithmic}[1]
%%   \STATE \(currbs \leftarrow [1]\);
%%   \FOR{($level=0; level<height; level++$)}
%%     \STATE \(prevbs \leftarrow currbs\);
%%     \STATE \(currbs \leftarrow {\bf level}(prevbs)\);
%%     \STATE ~
%%     \STATE \(currbs \leftarrow {\bf step}(T_l, currbs, prevbs)\);
%%     \IF{($currbs~=~[0~\ldots~0]$)}
%%       \RETURN $\emptyset$
%%     \ELSE
%%       \STATE \(currbs \leftarrow {\bf step}(T_r, currbs, prevbs)\);
%%       \IF{($currbs~=~[0~\ldots~0]$)}
%%         \RETURN $\emptyset$
%%       \ENDIF   
%%     \ENDIF 
%%   \ENDFOR
%%   \RETURN ${\bf leafs}(currbs)$
%%   \end{algorithmic}
%%   \caption{\label{a:tpj} {\bf Parallel-Traversing Join}($T_l$,$T_r$) returns all triples joining $T_l$ and $T_r$}
%% \end{algorithm}

\begin{figure}
  \centering
  \includegraphics*[bb=22 324 515 622,scale=0.35]{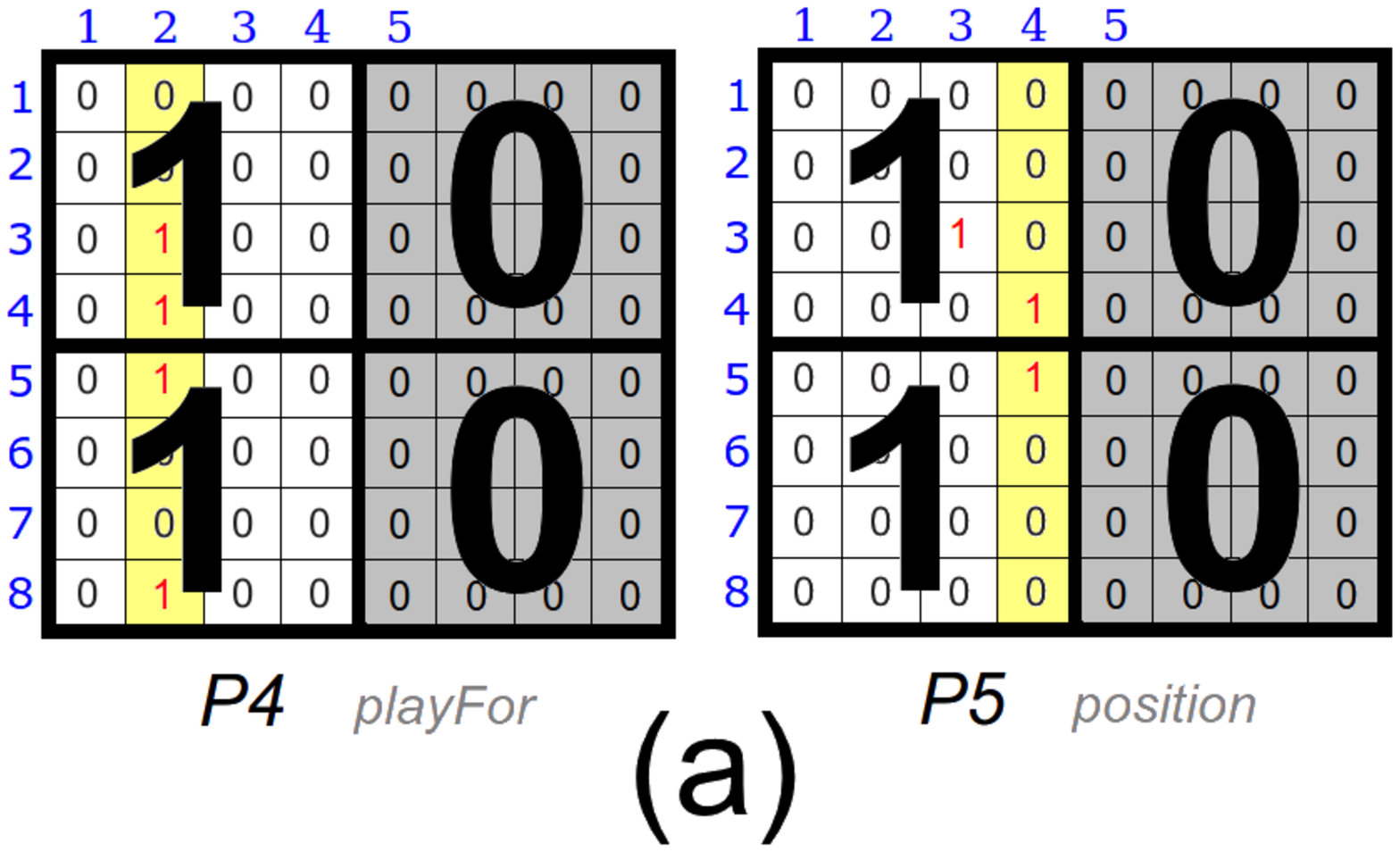}\\
  \includegraphics*[bb=22 324 515 622,scale=0.35]{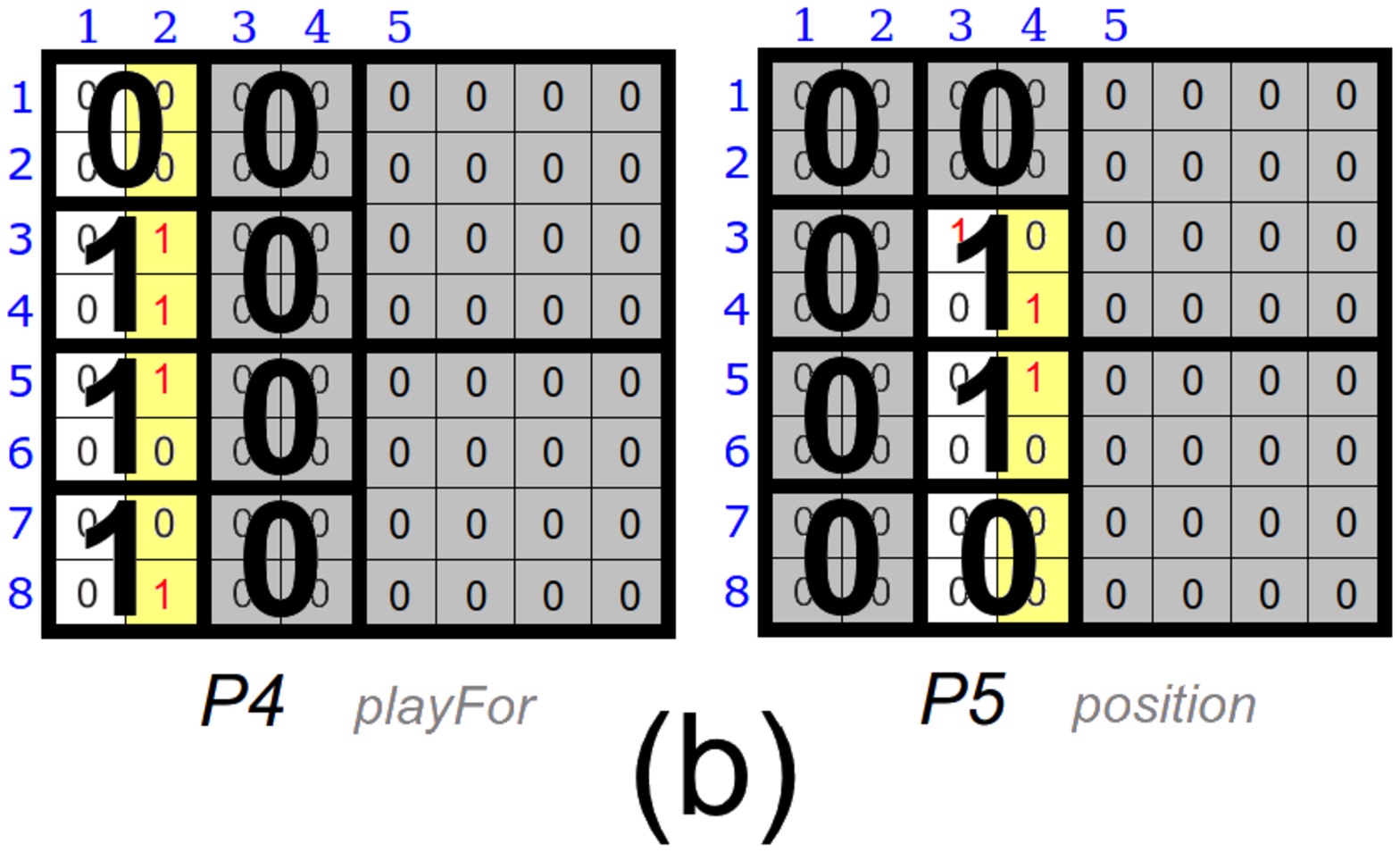}\\
  \includegraphics*[bb=22 324 515 622,scale=0.35]{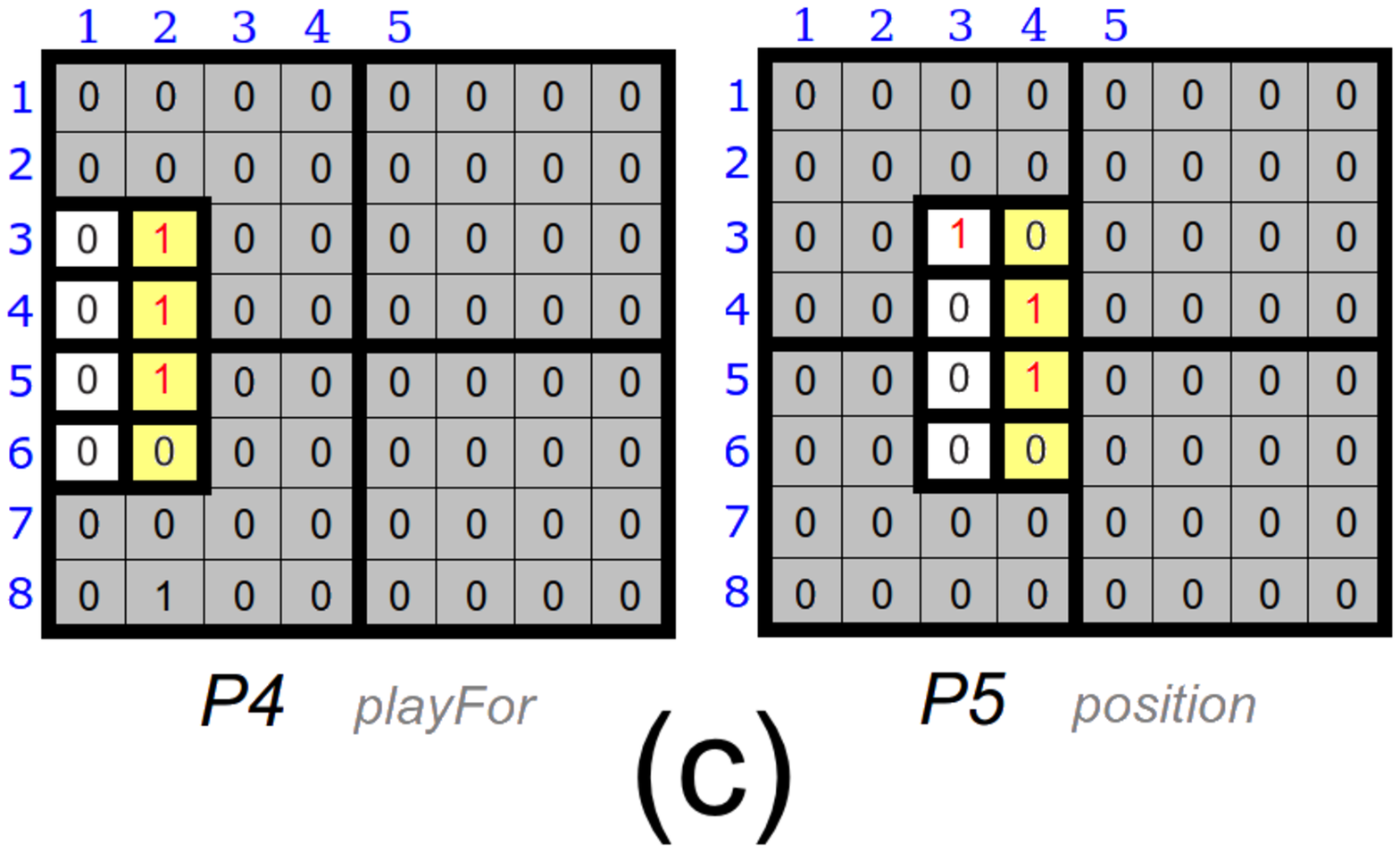}
  \caption{\label{fig:tpj} Example of interactive evaluation.}
\end{figure}

\begin{example}
Figure \ref{fig:tpj} illustrates how k$^2$-{\sc triples} implements the
interactive evaluation of the join query shown in Figure \ref{fig:exsparql}(b).
The original SPARQL query {\tt (?X, playFor, Spanish Team) 
(?X, position, midfielder)}, is rewritten as {\tt (?X, 4,2) (?X,5,4)} by 
performing the ID-based replacement of each term. Thus, the join must be
carried out on the k$^2$-trees that respectively model the predicates {\tt 4}
and {\tt 5}. Both k$^2$-trees are represented in Figure \ref{fig:tpj}(a)%
\footnote{The relation {\tt (8,2)} is added to {\tt P4} in order to provide
a more interesting example of the interactive evaluation algorithm.}.
Columns {\tt 2} and {\tt 4} are respectively remarked for the predicates {\tt 4}
and {\tt 5}, since those are the ones we have to join.
We consider $k=2$ and the join is implemented as follows:
%% ($\mathcal{M}_{i;[a,b][c,c]}$ refers the submatrix $[a,b] \times [c,d]$ within
%% the k$^2$-tree representing the $i-th$ predicate):

\begin{enumerate}[(a)]
  \item The two matrices $\mathcal{M}_4$ and $\mathcal{M}_5$ are queried. 
They are divided into $k^2=4$ submatrices (Figure \ref{fig:tpj}(a)). Both 
right submatrices in both $\mathcal{M}_4$ and $\mathcal{M}_5$ are discarded 
    because they do not overlap with the columns involved in the current query.
    The two pairs of left submatrices have value 1, so both 
    may contain results. Thus, we recursively consider the top-left and
the bottom-left submatrices of $\mathcal{M}_4$ and $\mathcal{M}_5$.
Note that we could have had to make more than one recursive call per 
submatrix, had we obtained more than one relevant top or bottom cell in
$\mathcal{M}_4$ and $\mathcal{M}_5$ (not in this case, where the columns
are specified).
  \item In the top-left submatrices (Figure \ref{fig:tpj}(b)) we discard in
turn the right subsubmatrices in $\mathcal{M}_4$ and the left subsubmatrices
in $\mathcal{M}_5$, because they do not intersect the query column. Further,
both top subsubmatrices are 0, so we need consider only, recursively, the
bottom-left subsubmatrix of $\mathcal{M}_4$ paired with the bottom-right
subsubmatrix of $\mathcal{M}_5$. Similarly, the top-left and top-right
subsubmatrices of $\mathcal{M}_4$ and $\mathcal{M}_5$ are recursively 
considered on the bottom-left submatrices. 

  \item The last recursion level (Figure \ref{fig:tpj}(c)) compares leaves
in $\mathcal{M}_4$ and $\mathcal{M}_5$.
    As in the previous step, we discard the cells that do not overlap
    with the query columns, and intersect the remaining ones. Three cells
    are possible results in $\mathcal{M}_4$: {\tt (3,2)}, {\tt (4,2)} and {\tt (5,2)}, so only 
    their corresponding counterparts must be evaluated in $\mathcal{M}_5$. 
Whereas the cell
    {\tt (3,4)} has value 0, the other two ones, {\tt (4,4)} and {\tt (5,4)}, 
    contain 1-bits. Thus, the {\tt 4} and {\tt 5} represent the subjects in the final 
    query result: {\tt Iniesta} ({\tt 4}) and {\tt Xavi} ({\tt 5}).
\end{enumerate}
\end{example}

\subsection{Implementing Joins over k$^2$-{\sc triples}}

This section details how k$^2$-{\sc triples} uses the proposed algorithms for
resolving all the join operations classified in Figure \ref{fig:joins}. We will 
refer to T$_l$ and T$_r$ as the first and second patterns involved in each
class of join. In general, interactive evaluation can be used uniformly on all
the cases, whereas chain and independent evaluation can also be used with
different strategies depending on the type of join. As a general rule of
thumb, chain evaluation is preferable over independent evaluation when the
outcome of one side of the join is expected to be much smaller than the other.
Interactive evaluation, instead, adapts automatically to perform in the best
way on each case. Finally, we remark that we will use indexes SP and OP
whenever possible to restrict the set of predicates to consider when the
predicate is variable (we will nevertheless remark this when their usage is
less obvious).

\paragraph{Joins with no variable predicates} As explained, in these classes of
joins both triple patterns provide their predicates. This ensures 
high performance for interactive resolution because only two parallel 
k$^2$-tree traversals must be performed. Chain and independent evaluation are
also possible, depending on the number of variable nodes involved in each class:

\begin{itemize}
  \item{\em Joins A.} It is the simplest class because only the join variable 
    is not provided. Chain evaluation is advantageous when one operand has much
    fewer results than the other. Otherwise, independent evaluation is better,
    as it leverages that both patterns return their results in sorted order.
  \item{\em Joins B.} This class leaves as variable a non-joined node. The 
    subject node of T$_l$ is variable in the example: {\tt (?S,P$_1$,?X) 
    (?X,P$_2$,O)}. Chain evaluation is well-suited for this class because it
    firstly resolves T$_r$, obtains all the values {\tt X$_i$} for {\tt ?X}, and
    finally replaces them in T$_l$. In this way, T$_l$ is trasformed into a group
    of patterns in which {\tt ?X} is replaced by each {\tt X$_i$} retrieved 
    from T$_r$. The final result comprises the union of all the results retrieved
    for the group of patterns obtained from T$_l$. Nevertheless, 
    independent evaluation also applies for this class. On the one hand, T$_r$ is
    	resolved through a {\em reverse neighbors} query, which returns its results in
    	order. On the other hand, a range query returns all results for T$_l$, which must
    	then be sorted by {\tt X}. The results of both operations are finally intersected
    	producing the join result set.
  \item{\em Joins C.} Both patterns in the join leave as variables their
    non-joined nodes. Chain resolution firstly resolves 
    the pattern containing the less frequent predicate (i.e., containing fewer
    {\tt (S,O)} pairs), extracts all its pairs, and all their distinct 
    {\tt X}$_i$ components are then replaced in the other pattern (note we
must remove duplicates in the {\tt X}$_i$ list before replacing each in T$_r$). Then all the
    objects found in T$_r$ for each {\tt X}$_i$ are matched with all the 
    subjects found in T$_l$ for the same {\tt X}$_i$. Alternatively, 
    independent evaluation generates all the pairs from both operands, sorted
    by the {\tt ?X} component in each case, and intersects the sorted lists.
\end{itemize}

\paragraph{Joins with one variable predicate} These classes 
comprise a triple pattern providing its predicate, and another that leaves
it variable. In this case, interactive resolution traverses, in parallel, 
the k$^2$-tree associated to the given predicate, and the $preds$ different 
k$^2$-trees involved in the other triple pattern resolution. In each recursive
step, only a subset of the $preds$ k$^2$-trees stay active for the corresponding
submatrix. Chain and independent evaluation strategies are also possible 
depending on the number of variable nodes involved in each operand:

\begin{itemize}
  \item{\em Joins D.} This class, like Joins A, provides the two non-joined 
    nodes but includes a variable predicate (say, that of T$_r$). In this case, chain evaluation 
    firstly resolves T$_l$, obtains all the values {\tt X$_i$} for {\tt ?X},
    and finally replaces them in T$_r$ for its resolution, which
    becomes a set of {\em access to single cell} queries. Independent
evaluation is also practical.
    First, T$_l$ is efficiently resolved with a {\em direct neighbors} query
and its results are retrieved in order. Second,
    T$_r$ performs $preds$ {\em inverse neighbor} queries to obtain the result
    set for {\tt (?X,?P,O)}, which must be adaptively sorted (for grouping
    the {\tt X}$_j$ values) before the final intersection. Note that not only
the OP index can be used to restrict the predicates of T$_r$ to those 
related to {\tt O}, but we can also restrict using SP to the union of all the
{\tt X}$_i$ values.
  \item{\em Joins E.} This class is split into two subclasses according to the
    pattern that contains the unbounded predicate and the variable non-joined 
    node. 
    \begin{itemize}
      \item{\em E.1.} Chain evaluation can choose between two strategies,
depending on which starts with the smaller set of candidates. On the one hand, T$_r$, which contains the unbounded predicate and 
 	provides the non-variable node, can be firstly resolved and its
results be
	adaptively sorted to remove duplicates. These results {\tt X}$_j$ for {\tt ?X} are
	then replaced in T$_l$ for its final resolution using {\em inverse
neighbor} queries. On the other hand, we could collect all the {\tt (S,X$_i$)}
pairs from {\tt P}$_1$, remove duplicates in {\tt X}$_i$, and run {\em access
to single cell} queries on all the qualifying k$^2$-trees for T$_r$.
Independet evaluation is also possible, much as done for Join D operations.
      \item{\em E.2.} Chain evaluation firstly resolves T$_r$ and all their 
	bindings {\tt X}$_j$ for {\tt ?X} are then used for replacement in T$_l$ (which is
        full-of-variables in this case) using $preds$ {\em inverse neighbor}
queries. 
    \end{itemize}
  \item{\em Joins F.} In this class, T$_l$ only provides the predicate, and 
    T$_r$ is full of variables. Chain evaluation firstly resolves T$_l$, 
    filters duplicate {\tt X}$_i$ values, and these are finally used
    for resolving T$_r$. This last step is restricted using index SP.
\end{itemize}

\paragraph{Joins with two variable predicates}  The triple patterns in this
class leave their two predicates as variables. This means that 
interactive resolution traverses in parallel all the different k$^2$-trees involved
in each pattern resolution. Chain and independent evaluation can proceed as
follows.

\begin{itemize}
  \item{\em Joins G.} This class provides the non-joined nodes and leaves the
    predicates as variables. Chain evaluation firstly resolves T$_l$, its
    bindings for {\tt ?X} are cleaned from duplicates, and 
    these are finally replaced in T$_r$ for its resolution.
    Independent evaluation is also suitable. It retrieves the
results for each pattern sorted by their {\tt ?X} component, and then
intersects the sorted lists.
  \item{\em Joins H.} In this case, T$_l$ is full of variables and T$_r$ binds
    the non-joined node. Chain evaluation firstly resolves T$_r$ and its 
    results, clean from duplicates, are used for T$_l$ resolution.
\end{itemize}

\begin{table}
  \caption{Summary of joins resolution in k$^2$-triples
    ($^*$ means that removing duplicates is required for join 
     resolution)\label{t:joins}}{
  \centering
  \scriptsize
  \begin{tabular}{|c|c|c|c|c|c|}
    \fullhline
    {\bf Join}&\multirow{2}{*}{\bf Example}&\multirow{2}{*}{\bf Chain}&{\bf Inde-}
	&\multicolumn{2}{c|}{\bf Interactive}\\
    {\bf Class}&&&{\bf pendent }&{\bf T$_l$}&{\bf T$_r$}\\
    \fullhline\fullhline
    \multirow{2}{*}{\bf A}&\multirow{2}{*}{\tt (S,P$_1$,?X)(?X,P$_2$,O)}&T$_l$ $\rightarrow$ T$_r$&
    	\multirow{2}{*}{$\surd$}&\multirow{2}{*}{Direct}&\multirow{2}{*}{Reverse}\\
    &&T$_r$ $\rightarrow$ T$_l$&&&\\
    \fullhline\fullhline
    {\bf B}&{\tt (?S,P$_1$,?X)(?X,P$_2$,O)}&T$_r$ $\rightarrow$ T$_l$&$\surd^*$&Range&Reverse\\
    \fullhline\fullhline
    \multirow{2}{*}{\bf C}&\multirow{2}{*}{\tt (?S,P$_1$,?X)(?X,P$_2$,?O)}&T$_l^*$ $\rightarrow$ T$_r$& 
    \multirow{2}{*}{$\surd^*$}&\multirow{2}{*}{Range}&\multirow{2}{*}{Range}\\
    &&T$_r^*$ $\rightarrow$ T$_l$&&&\\
    \fullhline\fullhline
    {\bf D}&{\tt (S,P$_1$,?X)(?X,?P$_2$,O)}&T$_l$ $\rightarrow$ T$_r$&$\surd^*$&Direct&Reverse ($\times preds$)\\
    \fullhline\fullhline
    \multirow{2}{*}{\bf E.1}&\multirow{2}{*}{\tt (?S,P$_1$,?X)(?X,?P$_2$,O)}& T$_l^*$ $\rightarrow$ T$_r$& 
    \multirow{2}{*}{$\surd^*$}&\multirow{2}{*}{Range}&\multirow{2}{*}{Reverse ($\times preds$)}\\
    &&T$_r^*$ $\rightarrow$ T$_l$&&&\\
    \fullhline
    {\bf E.2}&{\tt (?S,?P$_1$,?X)(?X,P$_2$,O)}&T$_r$ $\rightarrow$ T$_l$&&Range ($\times preds$)&Reverse\\
    \fullhline\fullhline
    {\bf F}&{\tt (?S,P$_1$,?X)(?X,?P$_2$,?O)}&T$_l^*$ $\rightarrow$T$_r$&&Range&Range ($\times preds$)\\
    \fullhline\fullhline
    \multirow{2}{*}{\bf G}&\multirow{2}{*}{\tt (S,?P$_1$,?X)(?X,?P$_2$,O)}&T$_l^*$ $\rightarrow$ T$_r$ 
    &\multirow{2}{*}{$\surd^*$}&\multirow{2}{*}{Direct ($\times preds$)}&\multirow{2}{*}{Reverse ($\times preds$)}\\
    &&T$_r^*$ $\rightarrow$ T$_l$&&&\\
    \fullhline\fullhline
    {\bf H}&{\tt (?S,?P$_1$,?X)(?X,?P$_2$,O)}&T$_r^*$ $\rightarrow$ T$_l$&&Range ($\times preds$)&Reverse ($\times preds$)\\
    \fullhline
  \end{tabular}
  }
\end{table}

\vspace{0.4cm}\noindent
Table \ref{t:joins} summarizes all presented choices for each class of join.
The first column indicates the class of
join and the second column illustrates a representative of the corresponding
class. 
Column {\em chain evaluation} describes
how this join strategy is carried out, that is, T$_l$ $\rightarrow$ T$_r$
means that T$_l$ is firstly executed and its results are used for T$_r$ 
resolution, and vice versa. 
Column
{\em independent evaluation} indicates the classes where this strategy can
be efficiently used. 
Finally, column {\em interactive evaluation}
indicates the k$^2$-tree operations interactively performed for resolving each
triple pattern in the join. We indicate with ``$\times preds$'' the cases
where interactive operations involve unbounded predicates.

%% file: experimentation.tex
\section{Experimentation}
\label{s:exper}

This section studies the performance of k$^2$-{\sc triples} on a
heterogeneous experimental setup comprising real-world RDF datasets from
different areas of knowledge. We study both compression effectiveness and
querying performance, and compare these results with respect to a consistent
set of techniques from the state of the art.

%% MIRAR COMO ESTÁ HECHA LA EXPERIMENTACIÓN DE HEXASTORE (GRÁFICAS)

\subsection{Experimental Setup}

We run experiments on an
AMD-Phenom\texttrademark-II X4  955@3.2 GHz, quad-core (4 cores - 4 siblings: 1
thread per core), 8GB DDR2@800MHz, running Ubuntu 9.10. We built two prototypes:

\begin{itemize}
  \item k$^2$-{\sc triples}, the vertical partitioning on k$^2$-trees without
the SP and OP indexes.
  \item k$^2$-{\sc triples}$^+$, which enhances the basic vertical partitioning 
    model with the indexes SP and OP.
\end{itemize}

Both prototypes were developed in C, and compiled using {\tt gcc} (version 4.4.1) with
optimization {\tt -O9}.

\paragraph{RDF Stores} We compare our results with respect to three 
representative techniques in the state of the art (Section \ref{s:state}):

\begin{itemize}
  \item A {\em vertical partitioning} solution following the approach of
    \cite{AMMH:07}. We implement it over 
    {\tt MonetDB\footnote{{\tt http://www.monetdb.org/} (MonetDB Database Server v1.6 (Jul2012-SP2))}}
    because it achieves better performance than the original C-Store
    based solution \cite{SGKNM:08}.
  \item A {\em memory-based} system implemented over 
    {\tt Hexastore\footnote{Hexastore has been kindly provided by its 
    authors.}}.
  \item A {\em highly-efficient} store: 
    {\tt RDF3X\footnote{\tt http://code.google.com/p/rdf3x/}}, which was 
    recently reported as the fastest RDF store \cite{HAR:11}.

\end{itemize}

All these techniques had been tested following the configurations and 
parameterizations provided in their original sources. 

\paragraph{RDF Datasets} We design a heterogeneous RDF data test comprising
four datasets from different areas of knowledge. We use it for testing 
k$^2$-{\sc triples} with respect to different data distributions, showing that 
our approach is competitive in a general scenario. The chosen datasets are the
following:

\begin{itemize}
  \item {\tt jamendo}\footnote{\tt http://dbtune.org/jamendo/} is a repository 
    of Creative Commons licensed music.
  \item {\tt dblp}\footnote{\tt http://dblp.l3s.de/dblp++.php} provides information
    on Computer Science journals and proceedings.
  \item {\tt geonames}\footnote{\tt http://download.geonames.org/all-geonames-rdf.zip}
    is a geographical database covering all countries and containing a large
number of placenames.
  \item {\tt dbpedia}\footnote{\tt http://wiki.dbpedia.org/Downloads351} is the
    semantic evolution of Wikipedia, so it is an encyclopedic dataset. {\tt dbpedia} 
    is considered the ``nucleus for a Web of Data'' \cite{ABKLCI:07}.
\end{itemize}

\begin{table}
  \caption{Statistical dataset description\label{t:stats}}{
  \centering
  \small
  \begin{tabular}{|c|r|r|r|r|r|}
    \fullhline
    {\bf Dataset}&{\bf Size (MB)}&{\bf \# Triples}&{\bf \# Predicates}&{\bf \# Subjects}&{\bf \# Objects}\\
    \fullhline
    {\tt jamendo}  &    144.18 &   1{,}049{,}639 &     28 &    335{,}926 &
440{,}604 \\
    {\tt dblp}     & 7{,}580.99 &  46{,}597{,}620 &     27 &  2{,}840{,}639 &
19{,}639{,}731 \\
    {\tt geonames} & 12{,}347.70 & 112{,}235{,}492 &     26 &  8{,}147{,}136 &
41{,}111{,}569 \\
    {\tt dbpedia}  & 33{,}912.71 & 232{,}542{,}405 & 39{,}672 & 18{,}425{,}128
& 65{,}200{,}769 \\
    \fullhline
  \end{tabular}
  }
\end{table}

The main statistics of these datasets are described in Table \ref{t:stats}. 
Note that some of the datasets contained duplicated triples, which have been 
deleted, hence sizes have been updated consequently. As 
can be seen, different sizes are chosen for overall scalability measurements.
In addition, {\tt dbpedia} lets us analyze how 
k$^2$-{\sc triples} perform when the number of predicates increases. It is 
worth remembering that queries with unbounded predicate are poorly resolved
using traditional solutions based on vertical partitioning.

\paragraph{Queries} We design experiments focused
on demonstrating the retrieval ability of all RDF stores included in our 
setup. First, we run triple pattern queries to analyze basic lookup 
performance. These results feed back join experiments which, in turn, predict
the core performance for BGP resolution in SPARQL.

We design a testbed\footnote{The full testbed is available at {\tt http://dataweb.infor.uva.es/queries-k2triples.tgz}} of randomly generated queries covering the entire spectrum of triple patterns and joins. For each dataset, we consider 500 random triple patterns of each type. Note that in all datasets, except for {\tt dbpedia}, the triple pattern {\tt (?S,P,?O)} is limited by the number of different predicates.  

Join tests are generated by following the aforementioned classification (A-H) (as shown in Figure \ref{fig:joins} for Subject-Object joins), and for each one we obtain specific joins Subject-Object (SO), Subject-Subject (SS), and Object-Object (OO). We generate 500 random queries of each join and perform a big-small consideration according to the number of intermediate results: for each join we take the product of the number of results for the first triple pattern and the results of the second triple pattern in the join. Given the mean of this product, we randomly choose 25 queries with a number of intermediate results over the mean (joins {\tt big}) and other 25 queries with fewer results than the mean (joins {\tt small}).

We design two evaluation scenarios to analyze how I/O transactions penalize
on-disk RDF stores included in our setup. On the one hand, the {\em warm}
evaluation was designed to favor query results to be available in main memory. 
It was implemented taking the mean resolution time of six consecutie repetitions
of each query.
%% by running queries five times and taking the result of the following repetition (the sixth one). 
On the other hand, the 
{\em cold} evaluation illustrates a real scenario in which queries are 
independently performed. All the reported
times were averaged on five independent executions in which the elapsed time
was considered.

\subsection{Compression Results}

We firstly focus on compression performance to measure the ability of
k$^2$-{\sc triples} to work on severely reduced space. This
comparison involves on-disk based representations, {\tt MonetDB} and 
{\tt RDF3X}, and in-memory ones, {\tt Hexastore} and our two 
k$^2$-{\sc triples} based approaches. In these cases, we consider the
space required for operating the representations in main-memory. 
Table
\ref{t:compression} summarizes the space requirements for all stores and
datasets in the current setup. 

\begin{table}
  \caption{Space requirements (all sizes are expressed in MB)\label{t:compression}}{
  \small
  \begin{center}
  \begin{tabular}{|c|r|r||r||r|r|r|}
    \multicolumn{1}{c|}{}
     &\multicolumn{2}{c||}{\bf On-disk}
     &\multicolumn{1}{c||}{}
     &\multicolumn{3}{c|}{\bf In-memory}\\
    \cline{2-3}\cline{5-7}
    \multicolumn{1}{c|}{}
     &\multicolumn{1}{c|}{\tt MonetDB}
     &\multicolumn{1}{c||}{\tt RDF-3X}
     &\multicolumn{1}{c||}{}
     &\multicolumn{1}{c|}{\tt Hexastore}
     &\multicolumn{1}{c|}{k$^2$-{\sc triples}}
     &\multicolumn{1}{c|}{k$^2$-{\sc triples}$^+$}\\
    \cline{1-7}
%%    {\tt jamendo}  &    78.45 &   37.73 & 1{,}371.25 &   0.74 &    1.28 \\
%%    {\tt dblp}     &   448.69 &  260.84 & $\times$ &  82.48 &   99.24 \\
%%    {\tt geonames} &   900.91 & 3584.80 & $\times$ & 152.20 &  188.63 \\
%%    {\tt dbpedia}  &  1811.46 & 9757.58 & $\times$ & 931.44 & 1178.38 \\
    {\tt jamendo}  &     8.76 &    37.73 & &  1{,}371.25 &   0.74 &    1.28 \\
    {\tt dblp}     &   358.44 & 1{,}643.31 & & $\times$ &  82.48 &   99.24 \\
    {\tt geonames} &   859.66 & 3{,}584.80 & & $\times$ & 152.20 &  188.63 \\
    {\tt dbpedia}  & 1{,}811.74 & 9{,}757.58 & & $\times$ & 931.44 & 1178.38 \\
    \cline{1-7}
  \end{tabular}
    \end{center}
  }
\end{table}

Among previous RDF stores, {\tt MonetDB} is the most 
compact one. This is an expected result according to the compressibility 
chances of column-oriented representations \cite{AMF:06}. {\tt MonetDB} demands
roughly 4 times less space than {\tt RDF3X} for the smallest datasets, matching
the theoretically expected difference according to the features of each 
underlying model. This difference is greater for {\tt dbpedia}: in this 
case {\tt MonetDB} uses $\approx 5.4$ times less space than {\tt RDF3X}.
On the other hand, {\tt Hexastore} reports an oversized representation for {\tt jamendo} 
and cannot index the other datasets in our configuration. 

Nevertheless, as can be seen, k$^2$-{\sc triples} requires much less space 
on all the datasets. It sharply outperforms the other systems, taking 
advantage of its compact data structures. This result can be analyzed 
from three complementary perspectives:

\begin{itemize}
  \item k$^2$-{\sc triples} is more effective than column-oriented compression
    for vertically partitioned representations. The comparison between our 
    approach and {\tt MonetDB} shows that k$^2$-{\sc triples} requires 
    several times less space than that used by the column-oriented database.
    The space used by {\tt MonetDB} for the largest datasets is around 
    $2-5.5$ times larger than k$^2$-{\sc triples} and $1.5-4.5$ times larger
than k$^2$-{\sc triples}$^+$. Besides, we also provide indexed access by object
    within this smaller space.
  \item k$^2$-{\sc triples} allows many more triples to be managed in main
    memory. If we divide the number of triples in {\tt jamendo} ($1{,}049{,}644$)
    by the space required for their in-memory representation in 
    {\tt Hexastore} ($1{,}371.25$ MB), we obtain that it represents roughly $765$ 
    triples/MB. This same analysis, in our approaches, reports that 
    k$^2$-{\sc triples} manages almost $1{,}5$ million triples/MB, and 
    k$^2$-{\sc triples}$^+$ represents more than $800{,}000$ triples/MB. Although this
    rate strongly depends on the dataset, its lowest values (reported for
    {\tt dbpedia}) are $\approx 200{,}000$ triples/MB. This means that
    k$^2$-{\sc triples} increases by more than two orders of magnitude the number
    of triples that can be managed in main memory on {\tt Hexastore}.
  \item k$^2$-{\sc triples} provides full RDF indexing in a space significantly
    smaller than that used for systems based on sextuple indexing. This 
    difference also depends on the dataset; for instance, {\tt RDF3X} uses
    roughly $8-10$ times the space required by our techniques for representing
    {\tt dbpedia}.
\end{itemize}

Finally, we focus on the additional space required by k$^2$-{\sc triples}$^+$ 
over the original k$^2$-{\sc triples} representation. Leaving aside 
{\tt jamendo}, whose size is tiny in comparison to the other datasets, this 
extra cost ranges from $\approx 20\%$ for {\tt dblp} to $\approx 26.5\%$
for {\tt dbpedia}. Thus, the use of the additional SP and OP indexes incurs in an
acceptable space overhead considering that our representation remains the most
compressed one even adding these new indexes. Nevertheless, as explained below,
SP and OP indexes are mainly useful for datasets involving a large number of 
predicates.

\subsection{Querying Performance}
\label{ss:exp-q}

This section focuses on query time performance.
We report figures for the most prominent experiments within our setup.

\paragraph{Triple patterns} 
These initial experiments measure the 
capabilities of all stores for RDF retrieval through triple pattern 
resolution. These are the atomic SPARQL queries, and are 
massively used in practice \cite{AFMP:11b}.

\begin{figure}
  \centering
  \includegraphics*[scale=0.42]{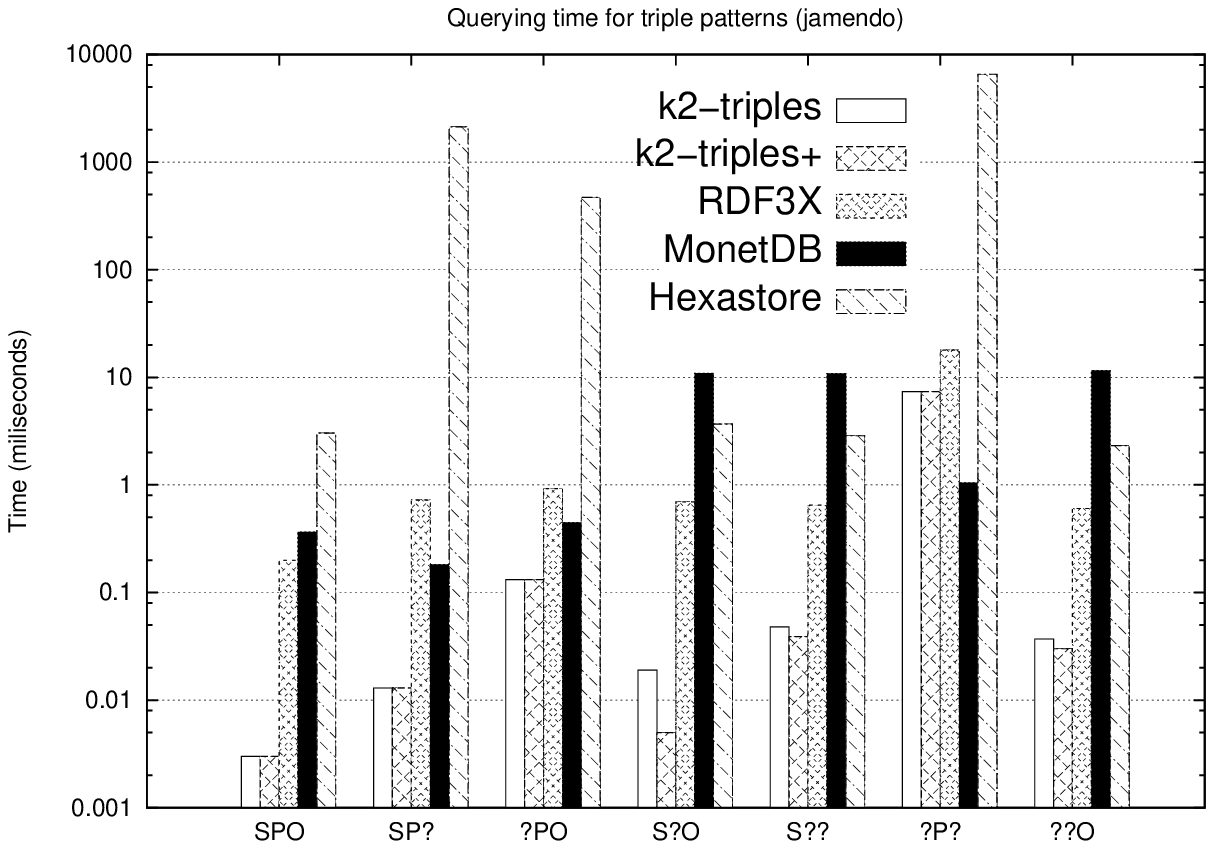}
  \includegraphics*[scale=0.42]{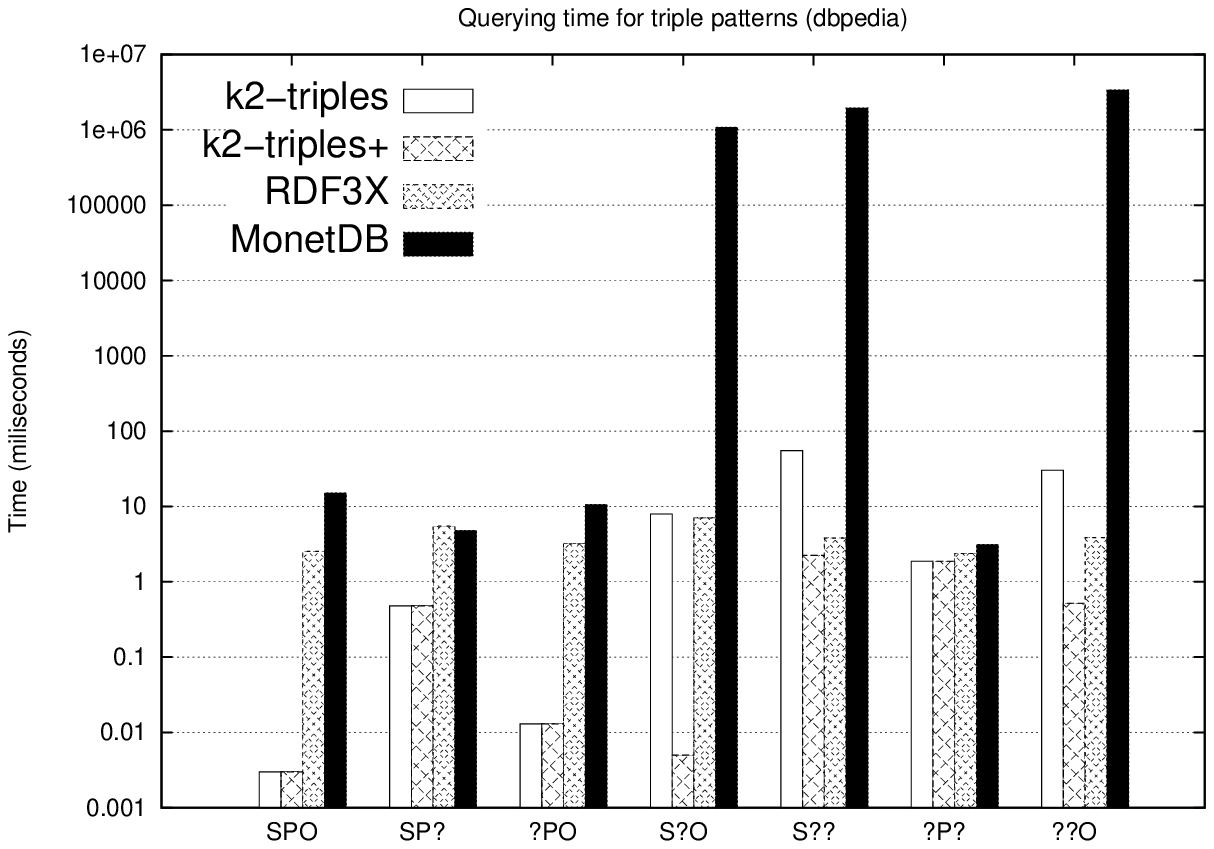}
  \caption{\label{f:triples} Resolution time (in milliseconds) for triple patterns
    in {\tt jamendo} and {\tt dbpedia} (warm scenario).}
\end{figure}

Figure \ref{f:triples} compares these times for {\tt jamendo} (left) and 
{\tt dbpedia} (right) in the warm scenario, which is the most favorable for 
on-disk systems.
The x axis lists all the possible
triple patterns\footnote{The pattern {\tt (?,?,?)}, which returns all 
triples in the dataset, is excluded because it is rarely used in practice.}
and groups the results for each system; resolution times (in milliseconds) are
reported in the y axis (logarithmic scale).

The comparison for {\tt jamendo} includes {\tt Hexastore}. As can be seen, this 
is never the best choice and it only outperforms {\tt MonetDB} in patterns with 
unbounded predicate. According to these results, we discard it because of its 
lack competitivity in the current setup. On the contrary, 
k$^2$-{\sc triples}$^+$ appears as the most efficient choice, and only 
{\tt MonetDB} slightly outperforms it for {\tt (?,P,?)} all collections but
{\tt dbpedia}.
Thus, in general, our approach reports the best overall performance for RDF
retrieval. This can be analyzed in more detail:

\begin{table}
  \caption{Resolution time (in milliseconds) for the patten 
    {\tt (?,P,?)} on {\tt dbpedia} (warm scenario)\label{t:preds}}{
  \begin{center}
  \begin{tabular}{|c|r|r|r|}
    \multicolumn{1}{c|}{}&k$^2$-{\sc triples}$^+$&{\tt RDF3X}&{\tt MonetDB}\\
    \cline{1-4}
    {\tt small} &  0.09 &  2.53 & 3.77 \\
    {\tt big}   & 24.57 & 14.88 & 6.14 \\
    \cline{1-4}
  \end{tabular}
  \end{center}
  }
\end{table}

\begin{itemize}
  \item Our approach overcomes the main vertical partitioning drawback and
    provides high performance for resolving {\em patterns with 
    unbounded predicate}. This is studied on {\tt dbpedia} 
    because in these queries scalability is more seriously compromised due to
	the  large number of predicates. k$^2$-{\sc triples}$^+$ leads the scene, 
    whereas {\tt RDF3X} is close for {\tt (S,?,?)}, falls behind for 
    {\tt (?,?,O)}, and is more than 2 orders of magnitude slower for
    {\tt (S,?,O)}. As expected, a larger improvement is achieved with respect
    to our original k$^2$-{\sc triples} (between 1 and 3 orders of magnitude),
    whereas our achievement is more significant in comparison with 
    {\tt MonetDB}: the difference ranges from roughly 5 
    orders of magnitude in {\tt (S,?,?)} to 8 orders for 
    {\tt (S,?,O)}.
  \item {\tt MonetDB} excels above the other systems in resolving the pattern 
    {\tt (?,P,?)}, but this comparison changes in {\tt dbpedia}. 
	This owes to the fact that predicates tend to be uniformly used in the 
    other datasets, whereas in {\tt dbpedia} some predicates are overused and
    the remaining ones are scarcely used. Thus, the number of results to be 
    obtained differs strongly and it affects the performance. Table 
    \ref{t:preds} summarizes resolution times for predicates returning 
    {\em small} and  {\em big} result sets\footnote{We average
    the number of triples per predicate and consider as {\tt small} those 
    ones with less ocurrences than the mean, and {\tt big} those predicates
    involved in more triples than the mean. We take 75 queries {\tt (?S,P,?O)}
    for each group.}. As can be seen, 
    k$^2$-{\sc triples}$^+$ dominates for less used predicates, whereas 
    {\tt MonetDB} is better when more results are retrieved. Thus, the 
    optimized column-oriented representation provides the fastest resolution
    when the predicate is used in numerous triples, whereas 
    k$^2$-{\sc triples}$^+$ outperforms it for more restictive predicates.
\end{itemize}

Finally, it is worth noting that k$^2$-{\sc triples}$^+$ obtains a competitive
advantage over the original k$^2$-{\sc triples} for datasets involving many
predicates. In the other datasets, this improvement is lower and both
techniques achieve comparable performance, yet k$^2$-{\sc triples} uses
sligthly less space as reported above. Nevertheless, we will use
k$^2$-{\sc triples}$^+$ in all the remaining experiments.
 
\paragraph{Joins} 
After studying triple pattern performance, the next stage in our setup
focuses on join resolution. The following experiments (i) analyze how our
three different join evaluation algorithms perform: {\em chain}, {\em independent} and 
{\em interactive}, and (ii) compare them with respect to RDF3X and 
MonetDB. All these experiments are performed in the warm scenario in order to
avoid penalizing on-disk solutions.

Figure \ref{f:jdbpedia} summarizes join results for {\tt dbpedia}. 
This figure comprises 9 plots, one for each class of join
according to the classification described in Figure \ref{fig:joins}. Each plot comprises
three subsets of joins: {\em Subject-Object} (SO), {\em Subject-Subject} (SS), and 
{\em Object-Object} (OO) in the x axis. The left group considers joins generating a
{\em small} amount of intermediate results, whereas the right group gives equivalent results
for joins involving {\em big} intermediate result sets. Resolution times (in milliseconds)
are reported on the y axis (logarithmic scale); note that times over $10^7$ milliseconds are
discarded in all the experiments. We analyze results:

\begin{figure}
  \includegraphics*[scale=0.3165]{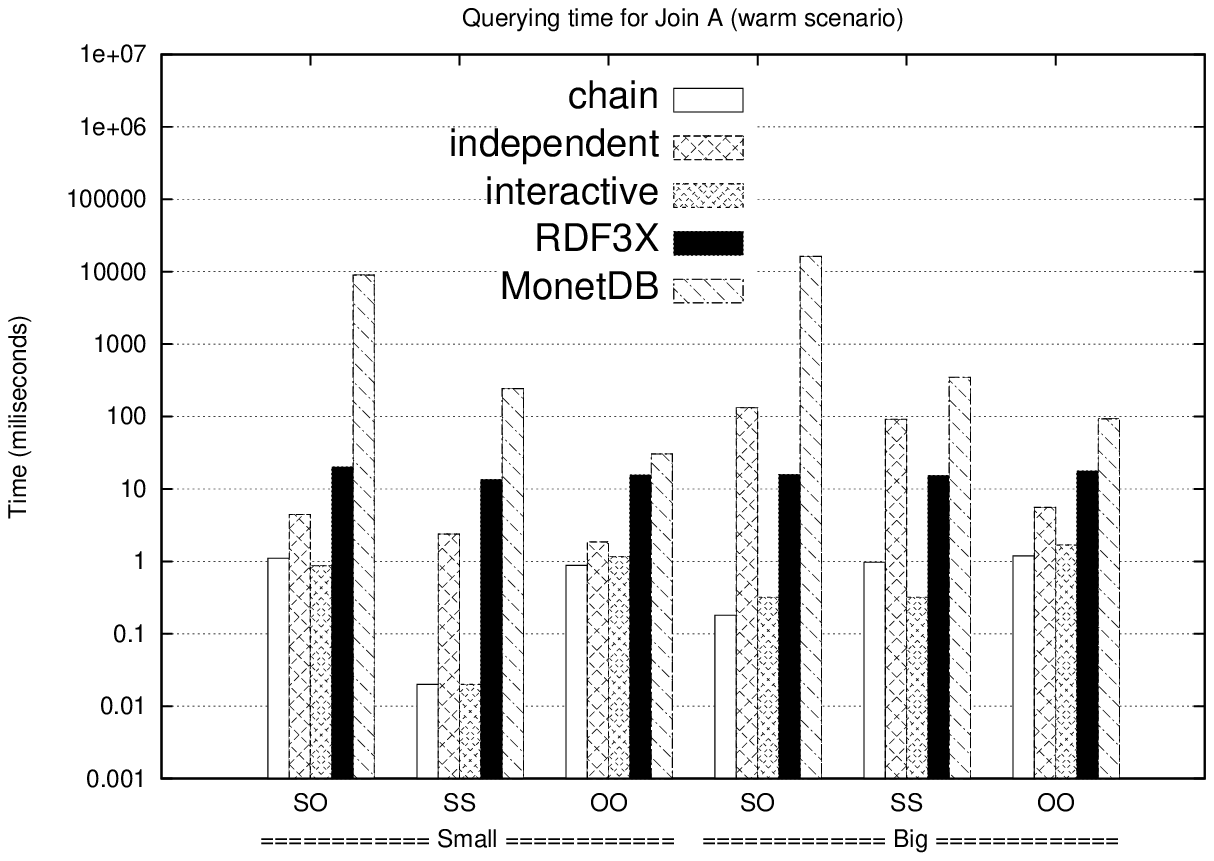}
  \includegraphics*[scale=0.3165]{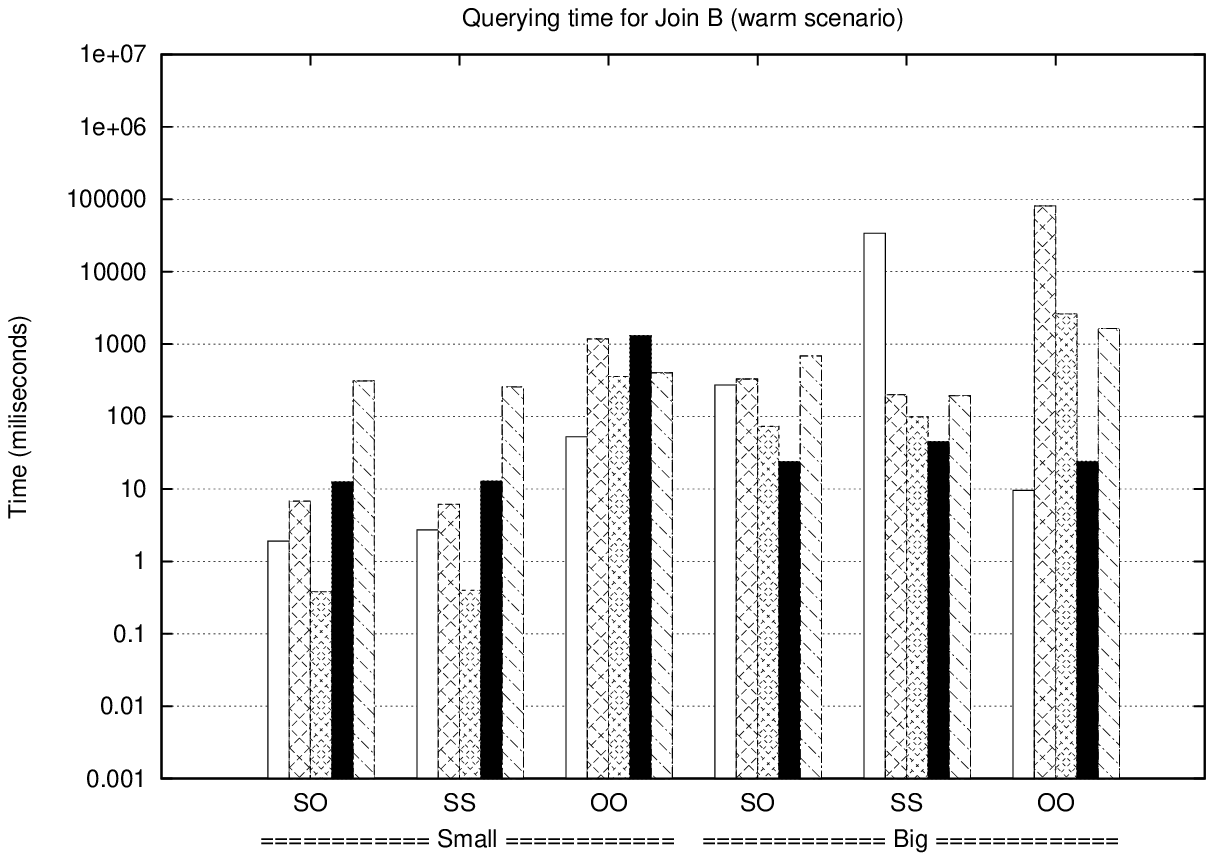}
  \includegraphics*[scale=0.3165]{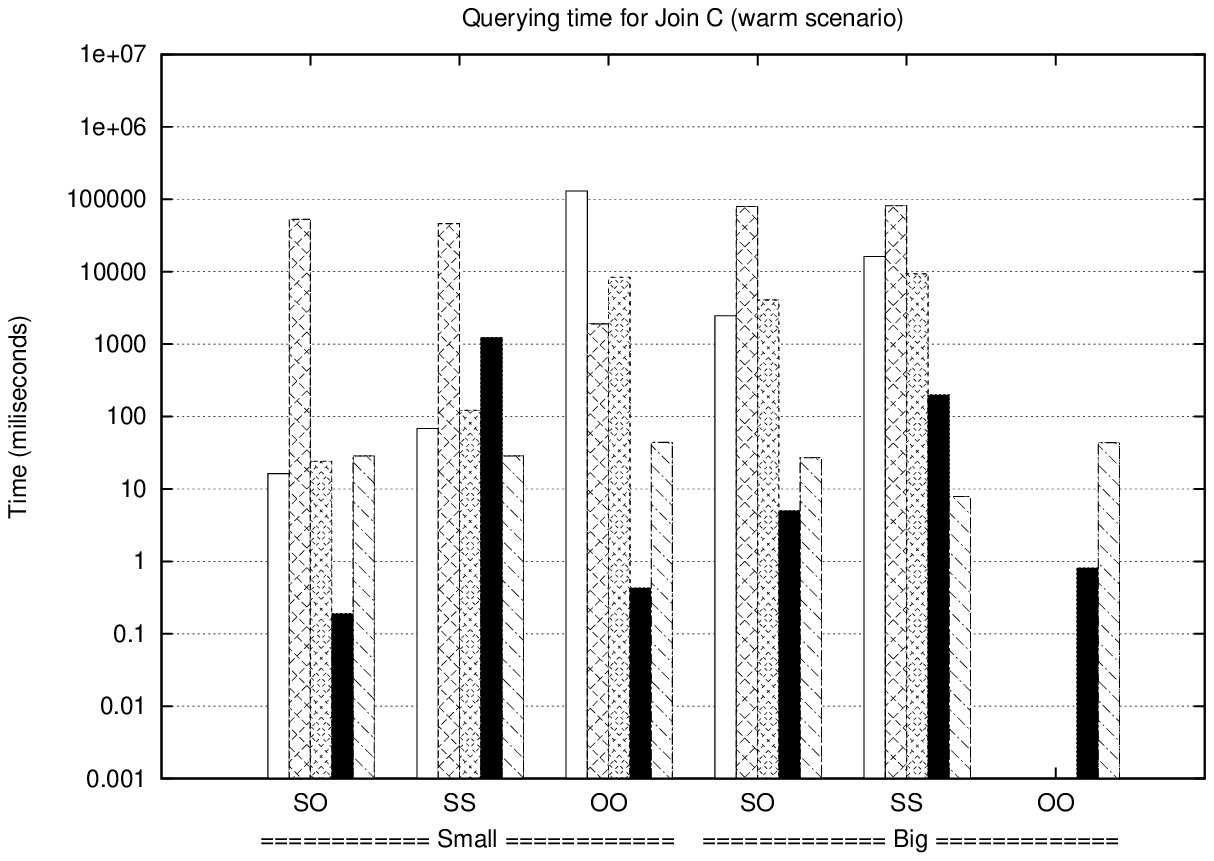}
  \\
  \includegraphics*[scale=0.3165]{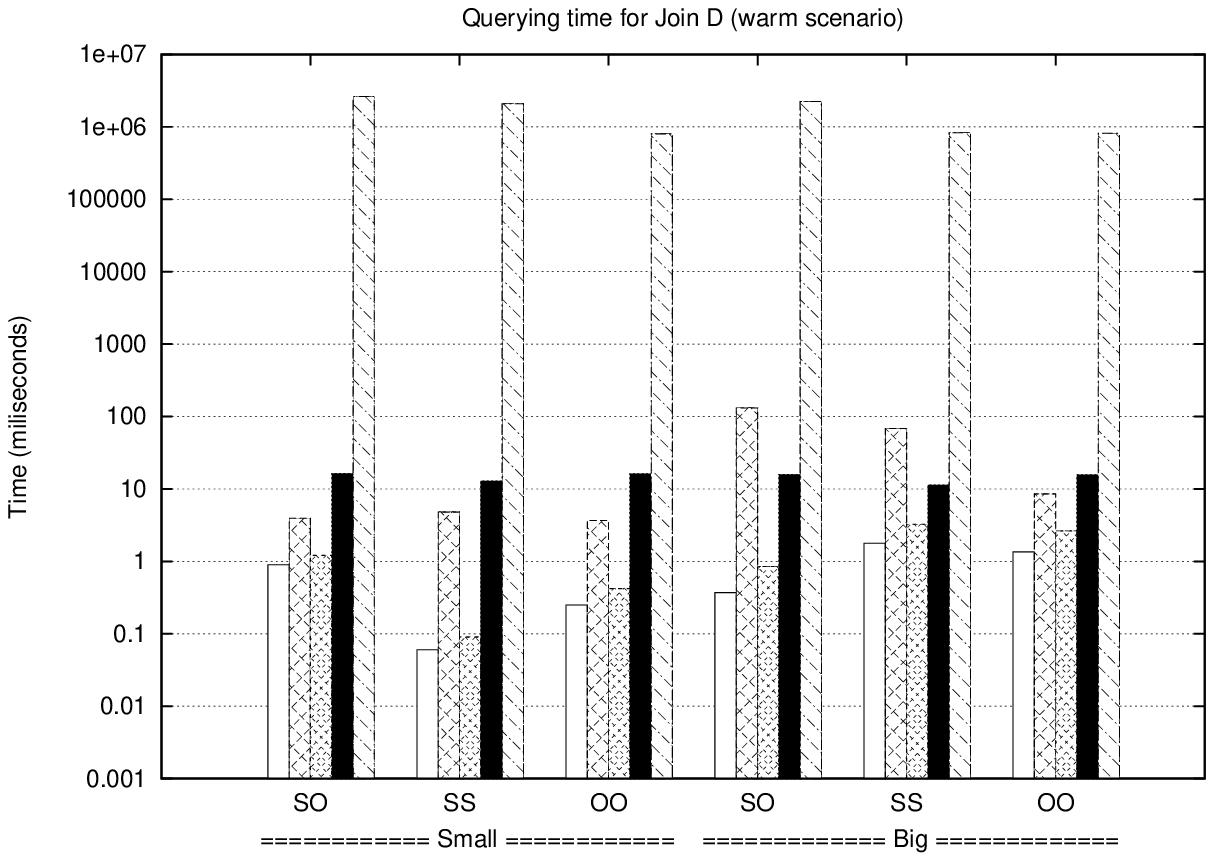}
  \includegraphics*[scale=0.3165]{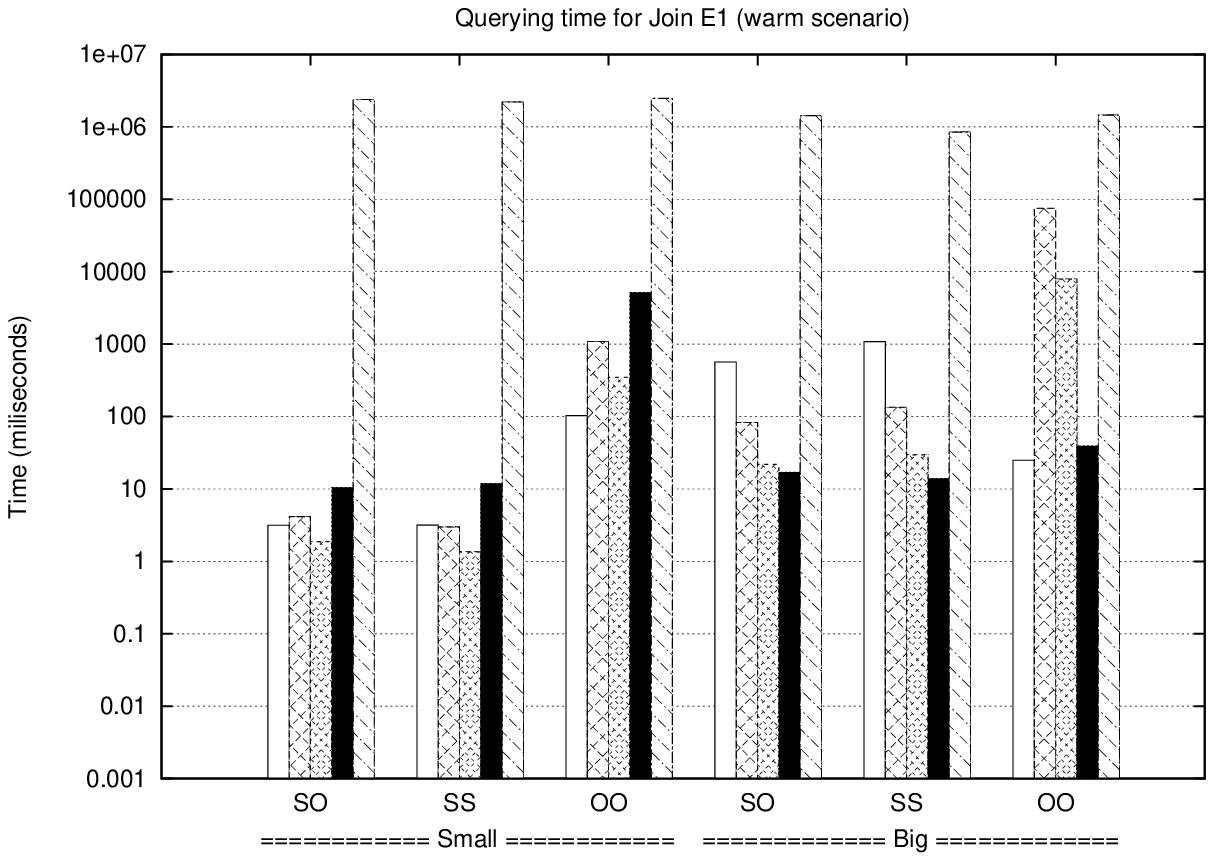}
  \includegraphics*[scale=0.3165]{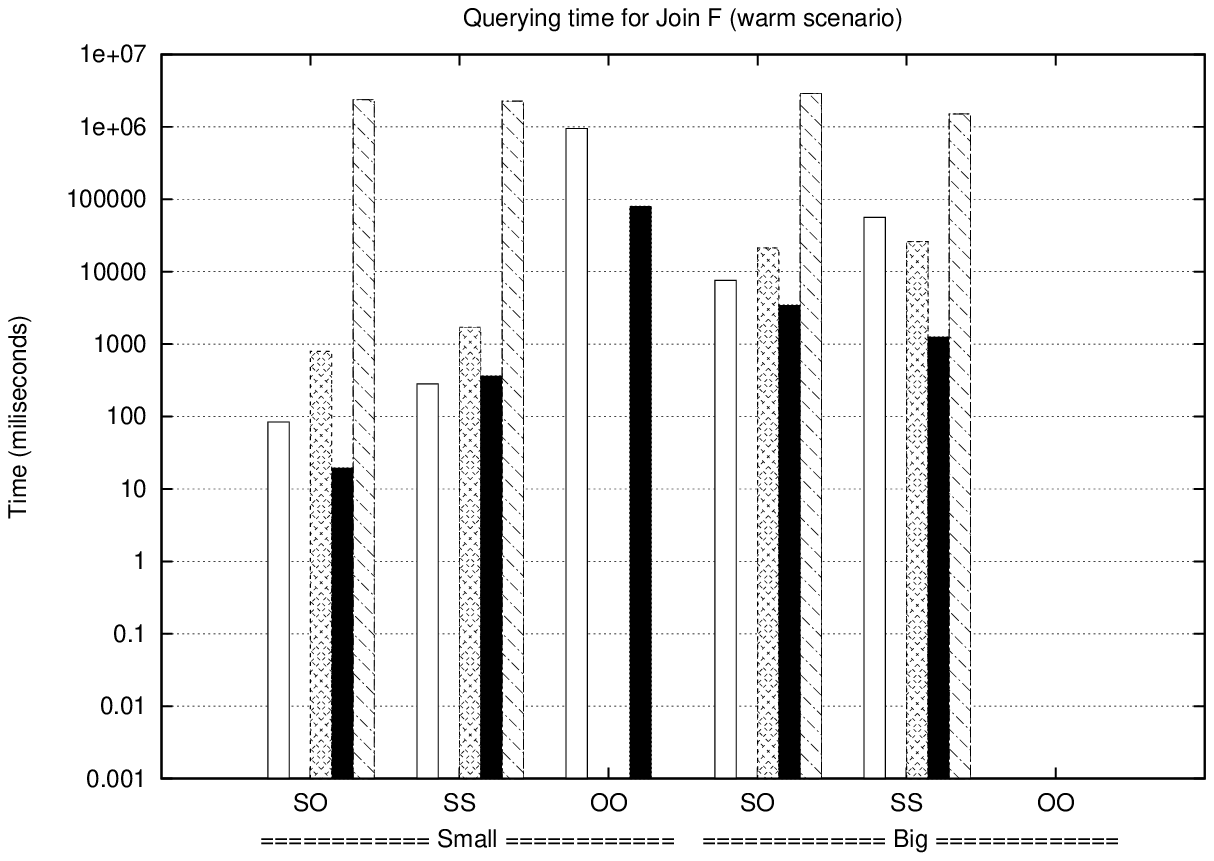}
  \\
  \includegraphics*[scale=0.523]{tests/joins/comparison/dbpedia/blank.eps}
  \includegraphics*[scale=0.3165]{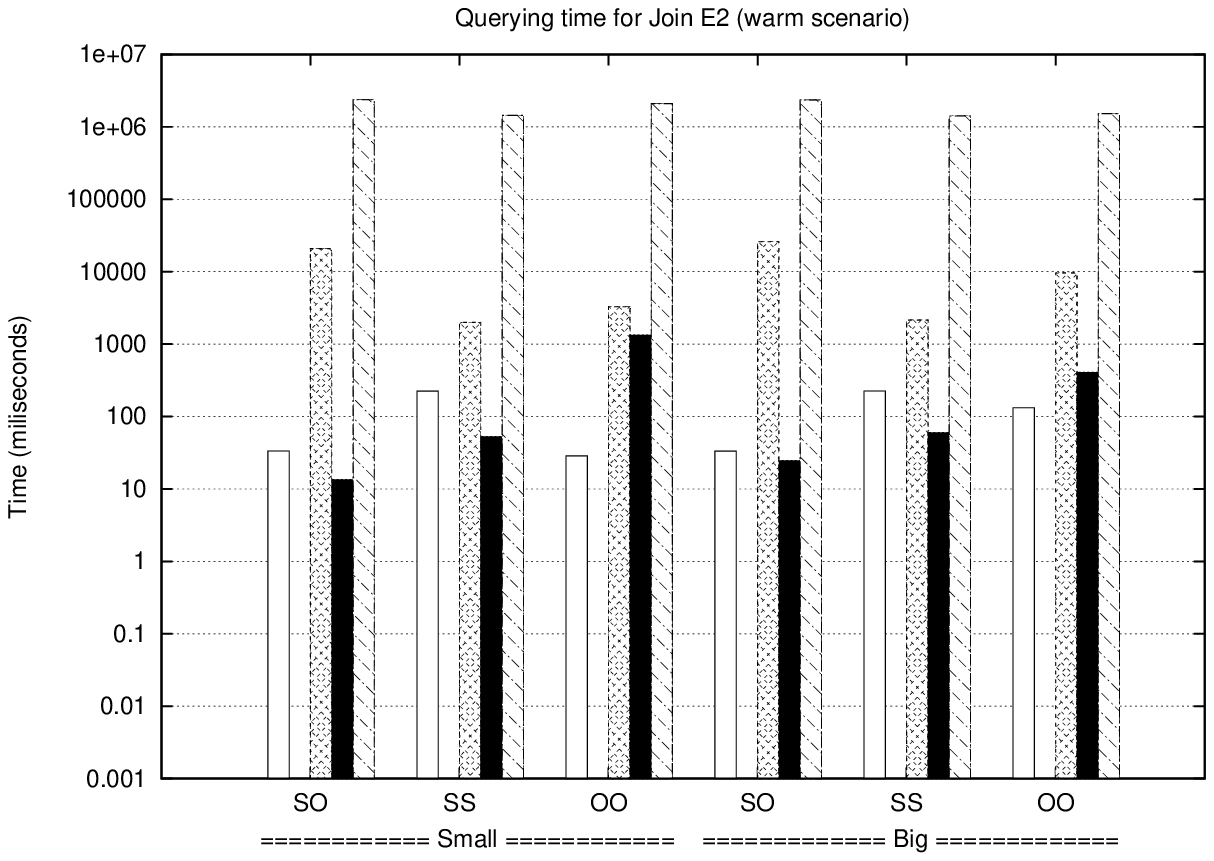}
  \\
  \includegraphics*[scale=0.3165]{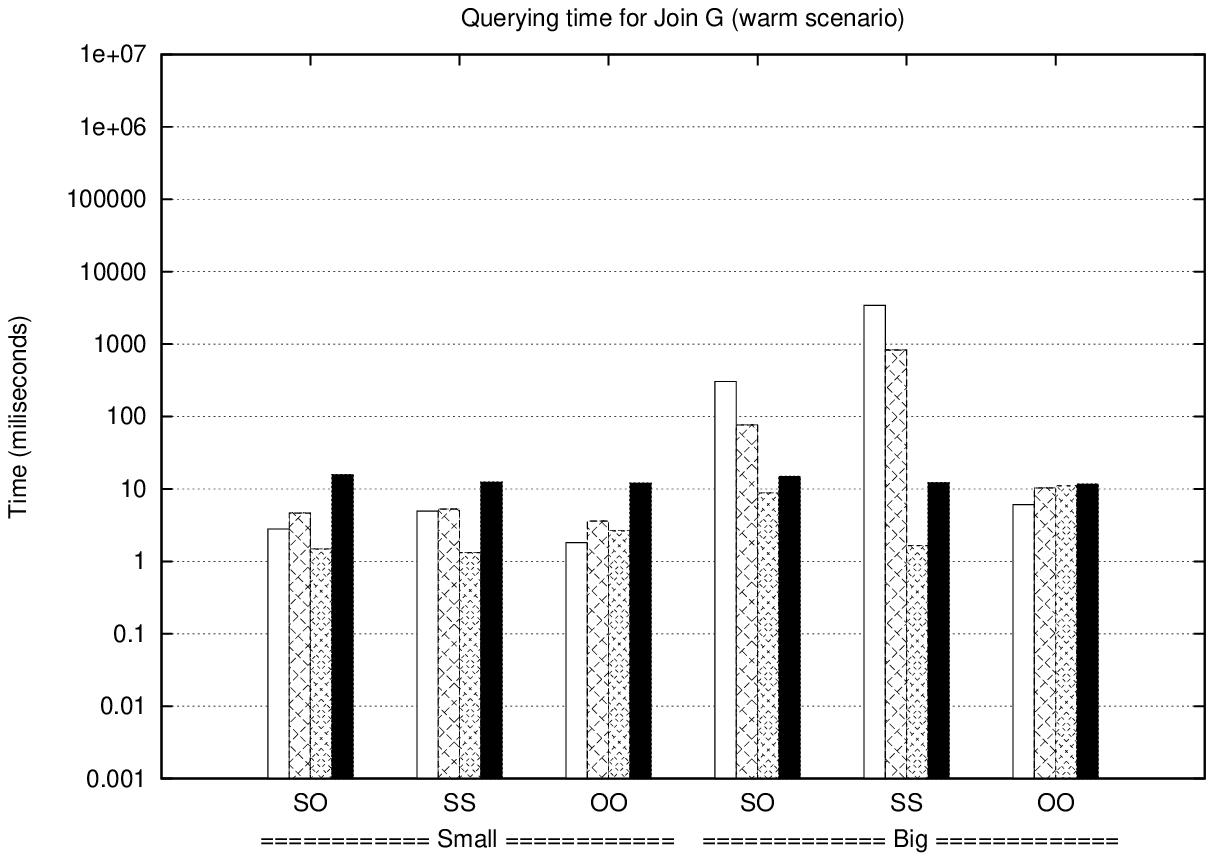}
  \includegraphics*[scale=0.3165]{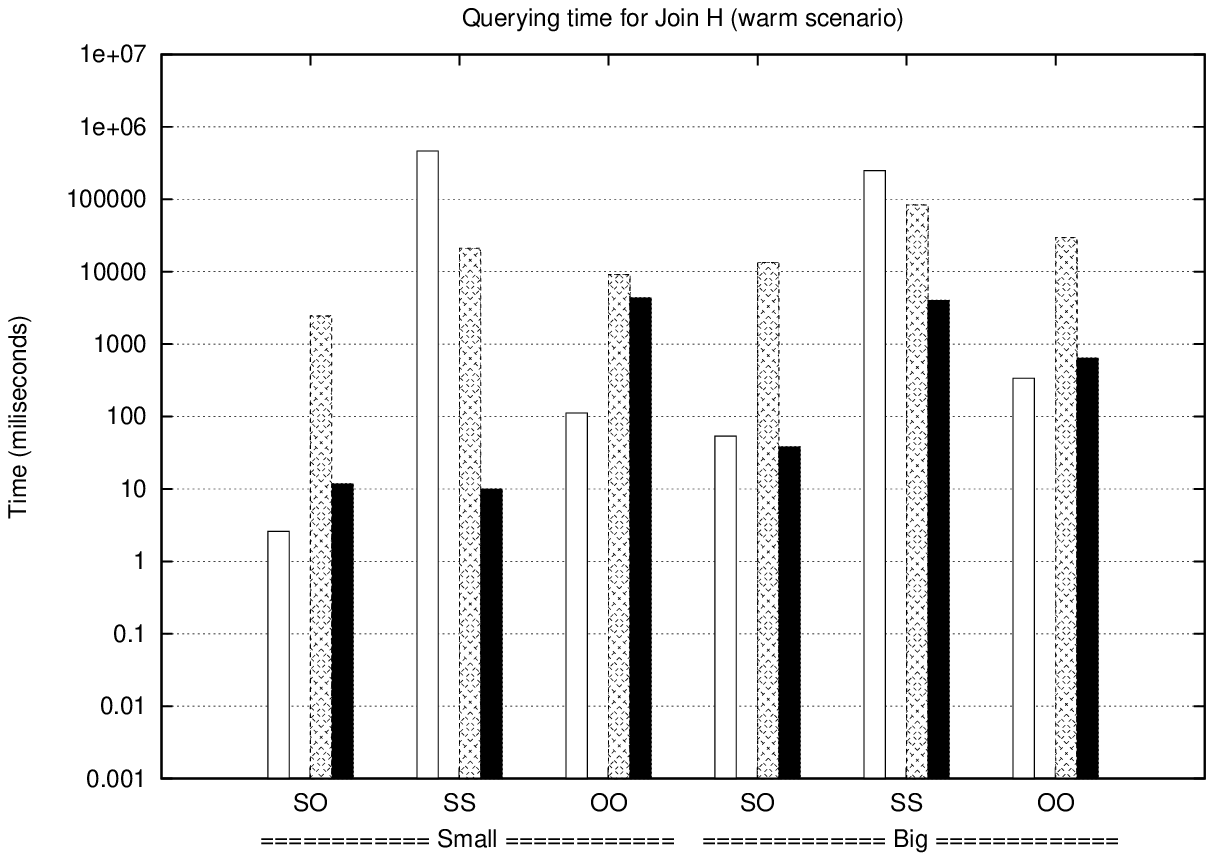}
  \vspace{-0.3cm}
  \caption{\label{f:jdbpedia}Resolution time (in milliseconds) for joins in {\tt dbpedia} (warm escenario). }
\end{figure}

\begin{itemize}
\item k$^2$-{\sc triples}$^+$ is the fastest technique for resolving joins in 
	which the value of the two not joined nodes are provided (classes A, D and
	G). This is mainly because all these classes are resolved using, 
	exclusively, direct and reverse neighbors queries, which are very efficient
	in practice. Both {\em chain} and {\em interactive} evaluation algorithms
	dominate {\bf Join A}: they report, at least, one order of magnitude of 
	improvement with respect to RDF3X and MonetDB. {\em Chain} evaluation is
	slightly faster in {\bf Join D}, improving upon RDF3X by more than one order
	of magnitude (except for OO big). Note that, in this case, MonetDB is no
	longer competitive since it pays the penalty of resolving a pattern with
	unbounded predicate. Finally, {\em interactive} is the fastest choice for 
	{\bf Join G}, although {\em chain} overcomes it for OO joins. While 
	k$^2$-{\sc triples}$^+$ is always faster than RDF3X in all cases,
	it is worth noting that 	differences are reduced due to the need of resolving 
	two patterns with unbounded predicate. Still we remain competitive, 
	while the performance of the vertical partitioning in MonetDB collapses
	(no times are drawn in this class).
\item The second column comprises joins leaving variable a not joined node and 
	fixing the other one (classes B, E, and H). k$^2$-{\sc triples}$^+$ and RDF3X
	share the lead in these experiments, whereas MonetDB remains competitive
	only in Join B, although it is never the best choice. On the one hand, the results
	for 	{\bf Join B} and {\bf E1} leads to similar conclusions. k$^2$-{\sc triples}$^+$ is
	the best choice for 	joins generating {\em small} intermediate result sets: 
	{\em chain} is fastest for OO, and {\em interactive} for SO and SS. RDF3X 
	overcomes k$^2$-{\sc triples}$^+$ when {\em big} intermediate result sets 
	are obtained, although our {\em chain} evaluation obtains the best
	perfomance for OO joins. On the other hand, {\bf Join E2} and {\bf H} give
	similar conclusions, as well. In this case, RDF3X always achieves the best times,
	except for OO joins, in which {\em chain} evaluation is the most efficient
	choice again. In this case, {\em interactive} evaluation is less competitive
	because it performs multiple range queries.
\item The third column comprises joins in which both triple patterns leave as 
	variables their not joined nodes (classes C and F). In {\bf Join C}, RDF3X
	is the best choice for SO and OO joins, whereas MonetDB wins for SS. Note
	that our approach remains competitive for SS and SO, but its performance is
	significantly degraded for OO. In {\bf Join F}, our {\em chain} 
	evaluation competes with RDF3X for the best times, overcoming it for SS 
	{\em small}. However, this arises as the most costly query; note that no 
	technique finishes on OO joins involving {\em big} intermediate results.
\end{itemize}

Summarizing, k$^2$-{\sc triples}$^+$ excels when triple patterns provide values
for the non-joined nodes, an it is clearly scalable when predicates are provided
as variables. Thus, in general terms, a query optimizer using 
k$^2$-{\sc triples}$^+$ must favor firstly joins A, D or G; then joins B, E, 
and H; and finally joins C and F. In any case, joins involving small 
intermediate result sets are always preferable over those generating big
intermediate results. 

These findings also apply, in general form, for the remaining datasets in our setup. 

%%% ESTO SE CAE???
%%% \subsection{Scalability Analysis}
%%% Aunque hayamos hecho toda la experimentaci\'on anterior, creo que esta
%%% prueba ser\'ia muy interesante de cara a mostrar hasta que n\'umero de 
%%% predicados nuestra propuesta es competitiva sin hacer ninguna otra cosa.
%%% Mi idea aqu\'i sería tomar particiones incrementales de DBPedia que
%%% vayan aumentando el n\'umero de predicados y, por consiguiente, el 
%%% n\'umero de triples. Con este podr\'iamos evaluar nuestra escalabilidad
%%% en espacio y tambi\'en en tiempo de consulta.
%%% }
%%% 
%%% {\bf Aumentar el n\'umero de predicados: primero el mas frecuente, despues el
%%%   segundo y as\'i... y probar con todas las consultas que tienen predicado
%%%   variable.}
%%% 
%%% 
%%% \paragraph{\bf Compression}
%%% 
%%% \paragraph{\bf Querying}
%%% 

%% file: conclusions.tex
\section{Conclusions and Future Work}
\label{s:conc}
This paper introduces a specific compressed index for RDF called 
k$^2$-{\sc triples}. Based on the well-known vertical partitioning model, this
technique represents (subject, object) pairs in very sparse binary matrices,
which are effectively indexed in compressed space using the novel k$^2$-tree 
stuctures. This modelling ensures the most compressed representations with
respect to a state of the art baseline and also provides the most efficient
performance for resolving triple patterns with fixed predicate. To overcome
the lack of scalability arising for patterns involving unbounded
predicates, two additional indexes are also represented in 
compressed space. This simple enhancement makes our technique 
dominant for most RDF retrieval activities. Moreover, we report
better numbers for join resolution, outperforming the state of the art for some
classes of join, while being competitive in most of the others.

Our future work focuses on getting a full-fledged RDF store over the current 
k$^2$-{\sc triples} approach. This means, on the one hand, designing a specific
query optimizer able to leverage the current retrieval features and also the 
reported performance for join operations. This would allow BGPs to be efficiently
resolved and set the basis for providing full SPARQL resolution. On the 
other hand, we will work to obtain a dynamic RDF storage that allows 
insertion, deletion and updating of triples over the current
data partitioning scheme. These objectives are strongly stimulated by the recent
advances reported on dynamic k$^2$-trees \cite{BBN:12}.